\documentclass[nofootinbib,
eqsecnum,tightenlines,11pt]{revtex4}

\usepackage{graphicx}
\usepackage{amsmath,amssymb,amsfonts,amsthm,stmaryrd,mathtools}
\usepackage{mathrsfs}
\usepackage{color}
\usepackage{multirow,bigdelim}
\usepackage{dsfont}
\allowdisplaybreaks[0]

\usepackage{axodraw4j}
\usepackage{pstricks}

\newcommand{\pic}[1]{
\vcenter{\hbox{\includegraphics[scale=0.6]{#1}}}
}

\def\be#1\ee{\begin{align}#1\end{align}}

\def\ba{\begin{eqnarray}}
\def\ea{\end{eqnarray}}

\def\f{\frac}

\def\q{\qquad}
\def\hs{\hspace{-0.3cm}}

\def\i{\mathrm{i}}
\def\SU{{\rm{SU}}}

\def\D{\mathcal{D}}

\begin{document}

\title{$(3+1)$--dimensional topological phases and self--dual quantum geometries encoded on Heegard surfaces}

\author{Bianca Dittrich}
\affiliation{Perimeter Institute for Theoretical Physics,\\ 31 Caroline Street North, Waterloo, Ontario, Canada N2L 2Y5}

\begin{abstract}
We apply the recently suggested strategy to lift  state spaces and operators for  $(2+1)$--dimensional topological quantum field theories to state spaces and operators for a $(3+1)$--dimensional TQFT with defects. We start from the $(2+1)$--dimensional Turaev--Viro theory and obtain a state space, consistent with the state space expected from the Crane--Yetter model with line defects. 

This work has important applications for quantum gravity as well as the theory of topological phases in $(3+1)$ dimensions. It provides a self--dual quantum geometry realization based on a vacuum state peaked on a homogeneously curved geometry. 
The state spaces and operators we construct here provide also an improved version of the Walker--Wang model, and simplify its analysis considerably. 

We in particular show that the fusion bases of the $(2+1)$--dimensional theory lead to  a rich set of bases for the $(3+1)$--dimensional theory. This includes a quantum deformed spin network basis, which in a loop quantum gravity context diagonalizes spatial geometry operators.  We also obtain a dual curvature basis, that diagonalizes the Walker--Wang Hamiltonian.  

Furthermore, the construction presented here can be generalized to provide state spaces for the recently introduced dichromatic four--dimensional manifold invariants.
\end{abstract}

\maketitle


\section{Introduction}

\noindent 
In this paper we will construct a state space and operators for a $(3+1)$--dimensional topological quantum field theory with line defects, based on the $(2+1)$--dimensional Turaev--Viro topological invariant \cite{TV}. There are interesting and timely applications for quantum gravity  as well as for the theory of topological phases in condensed matter. Additionally the techniques discussed here may allow a generalization to  recently introduced topological invariants for 4--dimensional manifolds, introduced by B\"arenz and Barrett \cite{BB4D}.   For the model discussed here we construct different bases, which generalize the fusion bases for the $(2+1)$--dimensional extended topological quantum field theories \cite{KKR,Wu,DG16}, and reveal a fascinating duality. These bases will allow a wide range of further applications, e.g. for background independent definitions of entanglement entropy  \cite{ABC16b} or for the construction of  coarse graining schemes \cite{ABC16a}. 
Let us shortly discuss these various points:\\
~\\
{\bf Quantum geometry realizations and quantum gravity with cosmological constant:} \\
For canonical  approaches to quantum gravity, such as loop quantum gravity, one is interested in a state space describing the kinematical geometric configurations of a $d$--dimensional hypersurface embedded into a $(d+1)$--dimensional manifold. Here one wishes in particular for a realization that respects (spatial) diffeomorphism symmetry. Such Hilbert spaces with diffeomorphism invariant states can be constructed from topological quantum field theories with defect excitations, as is suggested in \cite{TimeEvol}. The first such construction is known as Ashtekar--Lewandowski (--Isham) representation \cite{ALI} and involves a trivial TQFT, in which the vacuum is peaked on a  geometrically totally degenerate configuration. This fact makes the construction of states describing large scale geometries extremely complicated. This motivated the construction of a new quantum geometry realization \cite{DG} based on the BF topological field theory \cite{BF}, which involves a vacuum peaked on vanishing curvature instead. This vacuum solves indeed three-dimensional gravity without a cosmological constant. With the presence of a cosmological constant one would however expect a vacuum describing a homogeneously curved geometry. There are  a number of approaches to incorporate homogeneously curved geometry into the kinematical set--up of (loop) quantum gravity \cite{SmolinQ, BahrDittrichL, Girelli1}.   
An important aspect, making such a construction very attractive \cite{RV}, is that one expects the Hilbert space associated to a fixed triangulation\footnote{A continuum Hilbert space can be constructed via a so--called inductive limit, in which one considers a partially ordered set of arbitrarily fine triangulations. The resulting Hilbert space is expected to be infinite dimensional, which allows for operators with continuum spectrum.} to be finite dimensional. Thus the spectra of observables, which preserve the triangulation, are discrete.  This also avoids divergencies appearing in $\SU(2)$ based spin foam models \cite{SFDiv} and allows for (tensor network) coarse graining schemes \cite{TNWCG}.

In $(2+1)$ dimensions the Turaev--Viro (TV) \cite{TV} topological quantum field theory for ${\rm SU}(2)_{\rm k}$ describes Euclidean gravity with a cosmological constant. Thus one can expect that a Hilbert space based on the TV model with defect excitations leads to a suitable kinematical set--up for $(2+1)$--dimensional gravity with a cosmological constant. Such a Hilbert space, together with geometric operators, has been recently constructed in \cite{DG16}. Here braiding relations between strands of a graph defining the states, play a very important role. In generalizing this construction to $(3+1)$ dimensions one has to find a way to implement the braiding relations. 

This problem can be solved following a strategy suggested by Delcamp and the author \cite{3to4}:  To use a so--called Heegard splitting in order to represent  a three--dimensional (triangulated) manifold via a so--called Heegard surface.\footnote{See \cite{smolin2} for a much earlier suggestion to use surfaces to represent $(3+1)$--dimensional state spaces for quantum gravity.}  On this Heegard surface we can define the Hilbert space and operators of the $(2+1)$--dimensional TV topological field theory. Imposing certain constraints, we can interpret this Hilbert space  as a state space for a $(3+1)$--dimensional topological theory. This automatically incorporates a notion of defects, which are confined to the one--skeleton of the triangulation. 

We will find that the constraints reduce the Turaev--Viro state space to a state space of  the Witten--Reshetikhin--Turaev (WTR) TQFT \cite{WTR} on the Heegard surface.  Indeed the WRT TQFT arises as a boundary theory of the Crane Yetter invariant  \cite{Barrett4DObs}, and we consider here a boundary given by a spatial hypersurface.  

The WTR invariant can be considered as a quantization of three--dimensional Chern--Simons theory. This brings us to recent work by Haggard, Han, Kaminski and Riello \cite{HHKR}, which construct spin foam amplitudes for four--dimensional building blocks via the Chern--Simons theory defined on the boundary of these building blocks.  \cite{HHKR} also provides a semi--classical analysis of these amplitudes leading to a phase space describing blocks of homogeneously curved geometry. In this work we will provide a quantization of this phase space.  We will encounter more similarities with \cite{HHKR}: namely longitudinal and transversal Wilson loops (or holonomies in \cite{HHKR}) that take over the role of both holonomies and conjugated fluxes. `Longitudinal' and `transversal' are with respect to the boundary  of  a blow--up of the one--skeleton of the dual to the triangulation. Such a boundary does in fact define a Heegard surface.\footnote{One difference is that here we depict the Heegard surface as the boundary of the blow--up of the one--skeleton of the triangulation, as it is the one--skeleton that carries the curvature defects.}  Here we will not only provide a quantization of a phase space describing homogeneously curved building blocks but also reveal a deep duality: we will construct two different bases that diagonalize the transversal and longitudinal Wilson loops respectively.

~\\
{\bf $(3+1)$--dimensional topological phases and their excitations:}\\
The study of topological phases and their defect excitations has seen enormous progress in the last years. This applies in particular to $(2+1)$ dimensions. For the $(3+1)$--dimensional case less systematic results are available,  in particular  regarding the understanding of defect excitations.

The Walker--Wang model, which is defined for a cubical lattice  \cite{WW},  generalizes the $(2+1)$--dimensional string net models \cite{LevinWen} to $(3+1)$ dimensions. String net models describe a state space of graphs labelled by objects of a pre--modular fusion category.  The string net models do provide the state space for the Turaev--Viro models (based on the same fusion category) with defect excitations.  The Crane--Yetter (as well as Turaev--Viro) model is believed to describe BF theory with a cosmological constant term \cite{BaezL}.  This theory is also argued to give the effective description for the Walker--Wang model  \cite{Burnell}.  Thus one can conjecture that the Walker--Wang model describes the state space for the Crane--Yetter TQFT's with defect excitations.   

The Walker--Wang model is defined by a state space and a Hamiltonian. An analyses of the excitation content of these theories is provided by Keyserlingk et al \cite{Burnell}, which in particular shows that the ground state degeneracy and the properties of the excitations depend crucially on whether the fusion category is modular or not. 

Here we will provide an improved version of the Walker--Wang model. (We discuss modular fusion categories only, non--modular (but pre--modular) categories will be considered in \cite{BCtoappear}.) Our technique allows the definition of this model for arbitrary lattices and topologies. We furthermore provide different bases for the state space, in particular a basis that diagonalizes the (plaquette terms of the) Walker--Wang Hamiltonian. This makes the analyzes of the excitation content straightforward. In particular we can show uniqueness of the ground state (in the modular case) for arbitrary lattices and topologies.  Furthermore it seems possible to replace the Crane--Yetter invariant with the recently introduced dichromatic invariants \cite{BB4D}, which can lead to a more intricate excitation structure and degeneracy of the ground state. 

The model we discuss here can be seen as a quantum deformation of lattice gauge theory.  One can understand lattice gauge theory configurations in terms of defect excitations \cite{ABC16a}. The notion of these excitations depends on a choice of vacuum. The bases that we will construct here, make the excitation content explicit with respect to two vacua that are dual to each other. This can help to define coarse graining schemes as well as (background independent) notions of entanglement entropy \cite{ABC16b}.

~\\
{\bf Duality and Fourier transform:}\\
We will identify an interesting set of bases for the $(3+1)$--dimensional state space we will be constructing. All these bases are represented by a graph whose edges are labelled 
by ${\rm SU}(2)_{\rm k}$ irreducible representations. 

 There are in particular two bases that are dual to each other. These are associated to the one--skeleton of the triangulation and the one--skeleton of the dual complex respectively.  The latter basis coincides\footnote{in the sense that both basis arise from  graphs dual to a triangulations, with links labelled by ${\rm SU}(2)_{\rm k}$ representations.} with a (quantum deformed)  spin network basis \cite{RovelliSmolin}, which diagonalizes (spatial) geometry operators, such as the areas of the triangles.  The dual basis is associated to the one--skeleton of the triangulation and can be interpreted to encode the curvature (in excess of the homogeneous curvature) concentrated on this one--skeleton.  We will therefore refer to it as curvature basis. It generalizes the fusion basis for the $(2+1)$--dimensional case, as it encodes the excitations away from the vacuum peaked on homogeneous curvature. (All bases do descend from a $(2+1)$--dimensional fusion bases on the Heegard surface, the geometrical interpretation of these bases  differs however for the $(3+1)$--dimensional theory.)

The spin network basis diagonalizes Wilson loops along the boundary of the triangles of the triangulation, which represent exponentiated flux operators \cite{DG,HHKR}. The curvature basis diagonalizes Wilson loops around the edges of the triangulation and represent holonomy operators, measuring curvature.  The transformation between these two bases is therefore a generalized Fourier transform.  A related duality transform, involving however expectation values of geometric observables in three- and four--dimensional state sum models, has been discussed in \cite{BarrettObs,Barrett4DObs}.   Here we also find that the spectra of both kinds of Wilson loops coincide, revealing a deep self--duality. 

The two bases also define two different vacua: setting all free labels of the curvature basis to be trivial we conjecture that we obtain the Crane--Yetter (or quantum deformed BF) vacuum. Setting all labels to be trivial for the spin network basis we obtain a (quantum deformation of the) Ashtekar--Lewandowski vacuum, expressed on a fixed triangulation or dual graph.  Note however that a refinement of these states (in order to define the continuum limit) would require different embedding maps \cite{BD12b,BD14}. The entire construction rather assumed a quantum deformed BF vacuum and we can therefore expect that the operators we consider here are cylindrically consistent with respect to this vacuum. Thus it should be straightforward to construct a continuum Hilbert space based on the techniques in \cite{DG}. We leave the question whether one can construct also an Ashtekar--Lewandowski vacuum based continuum Hilbert space, based on the techniques presented here, for future research.

~\\
{\bf Outline of the paper:}\\
In the next section \ref{outline} we will outline the construction for the $(3+1)$--dimensional state spaces. This needs two pre--requisites. First the construction of state spaces for the $(2+1)$--dimensional Turaev--Viro theory, which we will review in section \ref{section:surfaceH}. Secondly some basics on Heegard splittings, that we provide in section \ref{Sec:Heegard}.  We then construct the state spaces, bases and operators for the $(3+1)$--dimensional theory in section \ref{Sec:3+1}. In the section \ref{examples} we discuss a number of examples, that is state spaces for different choices of triangulations and lattices. We will in particular consider the 3--torus with a cubical lattice, that underlies the Walker--Wang model.  We close with a discussion and outlook in section \ref{discussion}.

\section{Outline of the construction}\label{outline}

We will follow the strategy outlined in \cite{3to4} in order to construct the state space for a $(3+1)$--dimensional topological quantum field theory with defects from the state space for a $(2+1)$--dimensional topological field theory. 

The state spaces for the $(2+1)$--dimensional topological field theories are defined on (possibly punctured) surfaces $\Sigma$. Since we are dealing with a topological field theory the state space depends only on the topology of the surface.  For the class of topological field theories discussed here degrees of freedom are associated to (equivalence classes of) non--contractible cycles on the surface $\Sigma$. 

We now wish to consider a $(3+1)$--dimensional topological field theory, which also associates degrees of freedom to non--contractible cycles. On the three--dimensional equal time hypersurfaces ${\cal M}$ we will allow for line defects, and more generally for a defect graph. We define, that cycles cannot be deformed across the edges of a defect graph.  For a sufficiently nice defect graph, in particular, if the defect graph coincides with the one--skeleton of a triangulation, we can encode the information on the first fundamental group  by a so--called Heegard diagram. This is given by a (Heegard) surface, which can be obtained as the boundary of a blow--up of the defect graph. The Heegard surface is in addition equipped with a set of curves which are contractible in the enclosing manifold complement, but not contractible on the Heegard surface itself. We will refer to these curves as ${\cal C}_2$ curves. 

Thus we can adopt the state space of a $(2+1)$--dimensional TQFT for the Heegard surface. To make the interpretation as a state space for a $(3+1)$--dimensional TQFT with defects viable, we have to impose constraints that do not allow any excitations to be associated to the ${\cal C}_2$ curves. Thus these curves appear effectively as contractible curves.  

We will apply this strategy to the state space of the Turaev--Viro topological quantum field theory (TV TQFT). Here braiding relations play an important role, making the presence of the Heegard surface particularly useful. 

To proceed we will first review the state space and operators for the TV TQFT. We will then provide the necessary basics on Heegard diagrams and handle decompositions. Finally we will use these techniques to construct the state space for the $(3+1)$ dimensional TQFT with defects. 

We provide some background material on the quantum group $ {\rm SU}(2)_{\rm k}$ and a related graphical calculus in the appendix. More extensive treatments can be found in e.g.\cite{Qbackground,DG16}.

\section{Hilbert space for Turaev--Viro states on (closed) surfaces}\label{section:surfaceH}

We consider a closed genus $g$ surface and want to construct an associated Hilbert space spanned by solutions of the TV TQFT for this surface \cite{KKR,Kir,DG16}. (Being a solutions means that the states are in the image of the projector defined by the path integral, or equivalently satisfy the constraints associated to the TV theory.)

This Hilbert space will be spanned by states based on graphs, whose links are labeled (or coloured) by ${\rm SU}(2)_{\rm k}$ representations,  embedded into the surface $\Sigma$. We will impose equivalence relations on such graph states. These equivalence relations impose the flatness\footnote{Here `flatness' means a quantum deformed flatness, that is in gravitational context states satisfying $F\sim \Lambda e \wedge e$. In condensed matter literature the corresponding constraint is also known as plaquette constraint of stabilizer.}  constraints. These will ensure that we can isotopically deform the graph, so that degrees of freedom are only associated to non--contractible cycles.  Gau\ss~constraints, derived from  (a quantum deformed version of)  gauge invariance, are imposed through the coupling rules of ${\rm SU}(2)_{\rm k}$, which will restrict the allowed colourings of the graphs. These constraints ensure in particular that the graphs, on which the states are based, cannot have open ends (labelled with non-trivial representations). 

In the following we define the graphs, colourings and equivalence relations in more detail:\\~\\
{\bf Graphs:} We consider trivalent graphs embedded into the surface $\Sigma$. For ${\rm SU}(2)_{\rm k}$, which has self--dual representations, we do not need an orientation for the strands of the graphs. 

~\\
{\bf Colourings:}  We colour the strands of a given graph with irreducible admissible representations of ${\rm SU}(2)_{\rm k}$, that is with lables $j=0,1/2,\ldots, {\rm k}/2$. These labels also correspond to simple objects from the fusion category  associated to ${\rm SU}(2)_{\rm k}$. For each node we impose a coupling condition: the three representations meeting at a (trivalent) node need to include the trivial representation in their fusion product. (For ${\rm SU}(2)_{\rm k}$ the ordering in this fusion product does not influence the coupling conditions, which are detailed in (\ref{Admissibility}).) \\
~\\
{\bf Equivalence relations:} On the space of embedded coloured graphs we impose the following equivalence relations:
\begin{itemize}
\item Strands can be (isotopically) deformed:
\be\label{StrandDeformation}
\pic{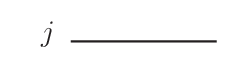}\,=\,\pic{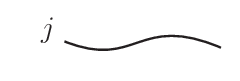}.
\ee
\item  Strands with trivial representations can be omitted:
\be
\pic{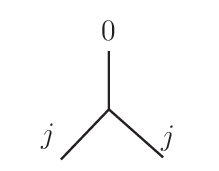}\,=\,\pic{Fig/HLine-j}.
\ee
\item The local connectivity of the graph can be changed by a so--called  $F$--move:
\ba\label{Fmove}
\pic{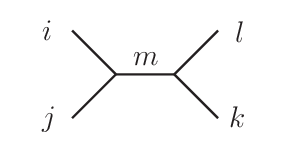}\,=\,\,\sum_nF^{ijm}_{kln}\pic{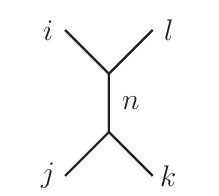} \q .
\ea
The $F$--symbol is defined in the appendix, equation (\ref{FDefinition}). 
\item Contractible loops of a graph can be annihilated using bubble moves:
\be\label{bubble move}
\pic{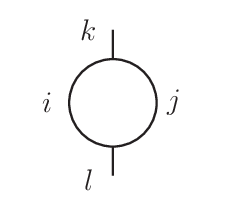}\,=\,\,\f{v_iv_j}{v_k}\delta_{kl}\delta_{ijk}\pic{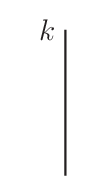}.
\ee
Here $v_j\,=\, (-1)^j \sqrt{d_j}$ is the square root of the quantum dimension of the representation $j$. With $d_j$ we denote the quantum number $[2j+1]$ defined in (\ref{qNumber}). Note the special case $k=l=0$ and thus $i=j$, stating that the $j$--bubble graph is equivalent to $v_j^2$ times the empty graph.
\end{itemize}

~\\
{\bf Crossings:}  We can also allow crossing of strands, but we need to keep track which strands are over--crossing and which are under--crossings. A crossing can be resolved into two three--valent nodes using the relations (\ref{resolution of crossing}). Note that  double over-- or under-- crossings can be also resolved by deforming one of the strands: 
\ba
\pic{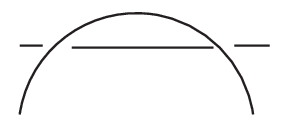}=\pic{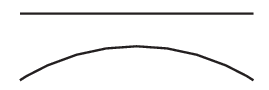}
\ea

A particular important identity involves a strand circling another strand:
\ba\label{smatrix1}
\pic{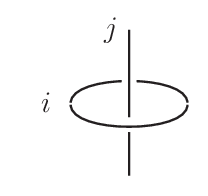}=\,\, \frac{s_{ij}}{s_{0j}}\pic{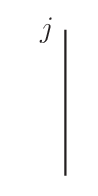},
\ea
where $s_{ij}$ is the so--called (rescaled) $S$--matrix defined in (\ref{SM1}, \ref{s matrix in terms of R}). 

~\\
{\bf Vacuum strands:}   Vacuum strands are defined as weighted sums over the strands labelled by admissible irreducible representations of ${\rm SU}(2)_{\rm k}$:
\be
\pic{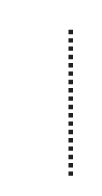} \coloneqq\,\, \frac{1}{\D} \sum_k v_k^2 \pic{Fig/VLine-k}.
\ee
A loop made out of a vacuum strand enjoys a special property known as sliding property. This holds for loops enclosing an arbitrary complicated region. The sliding property makes the region enclosed by the vacuum loop invisible to outside strands, in the sense that we can slide strands over the region:
\ba\label{sliding}
\pic{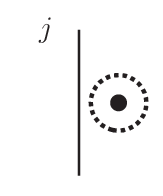}\,\,\,=\,\pic{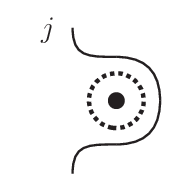}.
\ea
Thus we are allowed to deform Wilson lines, or holonomies operators, across the enclosed region. We can therefore interpret the vacuum loop as enforcing flatness over the enclosed region. 

Note that the insertion of a normalized vacuum loop, that is of a vacuum loop weighted with $1/{\cal D}$, defines a projection operators ${\cal P}$, satisfying ${\cal P}\circ {\cal P}={\cal P}$. To see this, slide one vacuum loop over the other loop and use the bubble move relation (\ref{bubble move}).

Furthermore, vacuum loops encircling a strand, force the associated representation label to be trivial: 
\ba\label{annihil}
\pic{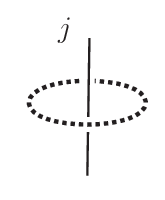}=\,\,\D\, \delta_{j0}.
\ea
This killing property holds (only) for modular fusion categories (such as ${\rm SU}(2)_{\rm k}$), in which the $S$--matrix is invertible. The killing property can be generalized to pre--modular fusion categories, but in this case only the non--transparent part of the strand going through the vacuum loop is annihilated, see \cite{BB4D}. Transparent objects are objects that braid trivially with all other objects of the category.  The killing property will play an important role in our discussion for the $(3+1)$--dimensional theory. 

~\\
{\bf Bases:} It is rather involved to find a set of independent states under the equivalence relations (\ref{Fmove}) and (\ref{bubble move}). However a systematic way of constructing a basis for a given genus $g$ surface is known \cite{kono,Koenig2,KKR}. It is a generalization of the so--called fusion basis for punctured spheres \cite{KKR}. These bases are defined to be orthonormal, which equips the Hilbert space with an inner product.

If the surface $\Sigma$ is a sphere, all cycles are contractible, and the equivalence relations (\ref{StrandDeformation}--\ref{bubble move}) can be used the reduce any graph to the empty graph. The Hilbert space associated to the sphere is therefore one--dimensional. The first non--trivial case is that $\Sigma$ has the topology of the torus. Two bases for the torus are depicted in equation (\ref{equTorusB}.)  The bases diagonalize over-- and under--crossing Wilson loops around the equator (for the basis depicted on the left side) or the meridian (for the basis depicted on the right side) respectively, as we will see further below. The two bases are connected by a unitary transformation
\ba\label{equTorusB}
\pic{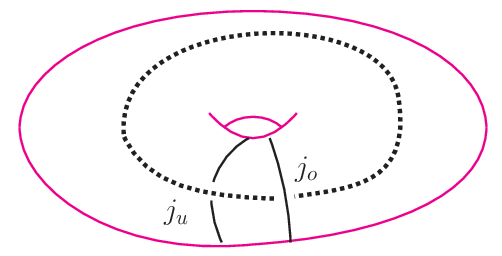} = \sum_{k_o,k_u}  \mathbb{S}_{j_oj_u,k_ok_u}     \pic{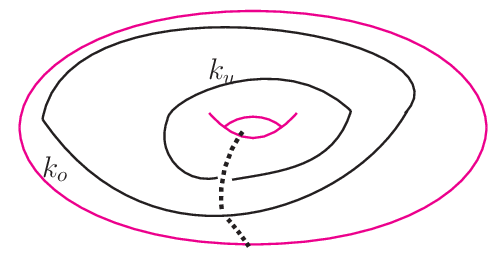}
\ea
 described by a so--called ${\mathbb S}$--matrix (here of the Drinfeld double of ${\rm SU}(2)_{\rm k}$), which factorizes into parts describing the over--crossing and under--crossing graphs respectively:
\ba
\mathbb{S}_{j_oj_u,k_ok_u}=S_{j_o k_o} \,\, S_{j_u k_u} \q .
\ea

\begin{figure}[!]
	\centering
	\begin{minipage}[b]{1\textwidth}
		\centering
		\includegraphics[scale = 0.7]{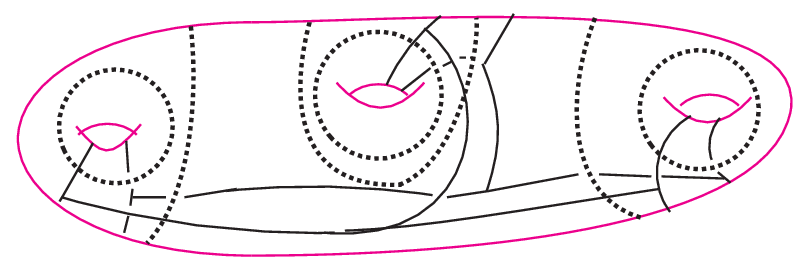}	
	\end{minipage}
	\caption{A basis for a genus 3 surface.  Three of the shown vacuum loops can be contracted to a trivial cycle after using repeatedly the sliding property with the remaining vacuum loops.   \label{Genus31}}
\end{figure}

For  a genus $g\geq2$ surface  we decompose the surface into pants, that is three--punctures spheres. To this end we need to cut the surface along $(3g-3)$ non--contractible curves. We will refer to this set of cutting curves as ${\cal C}_B$.
 We can construct a trivalent graph ${\cal F}$  dual to this set of curves.  (This graph is also called a spine.) That is  each link of ${\cal F}$ crosses one curve, and each pant component carries one node of the graph ${\cal F}$. A basis can now be constructed as follows:  We double the graph ${\cal F}$  to a double strand graph, where one copy ${\cal F}_o$ of the graph is formed from over--crossing strands and the other copy ${\cal F}_u$ from under--crossing  strands. Along each cutting curve we draw a vacuum loop, that over--crosses the under--crossing graph copy and under--crosses the over--crossing graph copy.\footnote{The set of vacuum loops will in general be over--complete, in the sense that vacuum loops can be slid over each other and then projected out. There will be however a minimal set of vacuum loops that cannot be further reduced. This corresponds to a set of independent cycles in the graph ${\cal F}$. Note however that for each redundant vacuum loop that we remove we need to multiply the state  with a factor of ${\cal D}$, so that the norm remains invariant.}  Figure \ref{Genus31} shows a choice for vacuum loops and the over-- and under--crossing graphs for a genus 3 surface.

The set of these states given by all admissible colourings of the double graph,  defines an orthonormal basis for the Hilbert space ${\cal H}(\Sigma)$.

Different bases, with different underlying choices for the spine ${\cal F}$ can be transformed into each other by two transformations:
\begin{itemize}
\item The $\mathbb{S}$ moves apply when two boundaries of a three--punctured sphere are glued to each other:
\ba\label{SStrafo}
\pic{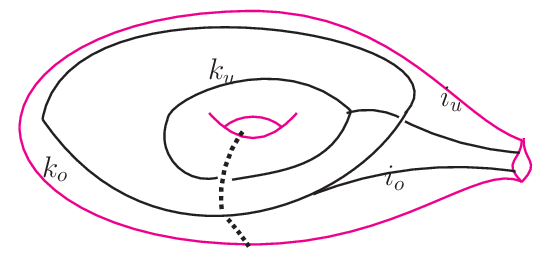}=\sum_{j_o,j_u}  \mathbb{S}^{i_oi_u}_{k_ok_u,j_oj_u} \pic{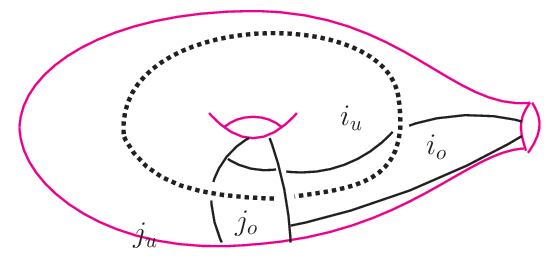} \q .
\ea
The $\mathbb{S}$ transformation factorizes again in an under--crossing and over--crossing part
\ba\label{SAB}
 \mathbb{S}^{i_oi_u}_{k_ok_u,j_oj_u}\,=\, A^{i_o}_{k_oj_o} \, B^{i_u}_{k_uj_u},
  \ea
 where the tensors $A$ and $B$ are defined in (\ref{equAB}).

\item Flip moves or $\mathbb{F}$ moves apply to a gluing of two three--punctured spheres to a four--punctured sphere:
\ba\label{FFmove}
\pic{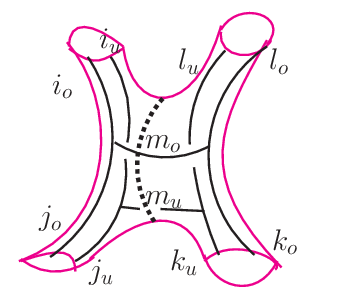}\,=\, \sum_{n_o,n_u} F^{i_oj_om_o}_{k_ol_on_o} \, F^{i_uj_um_u}_{k_ul_un_u}  \pic{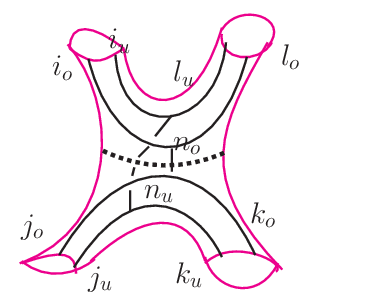} \q .
\ea
This transformation property follows from the $F$--move equivalence (\ref{Fmove}) and the fact that the vacuum loops around several punctures can be generated from the vacuum loops around the punctures itself. That is the vacuum loops around the punctured spheres in (\ref{FFmove}) are redundant, given that there are vacuum loops around the punctures itself (which are not depicted in  (\ref{FFmove})).
\end{itemize}
The fact that these transformations keep the splitting into an over--crossing and an under--crossing graph intact, will play an important role later-on.

~\\
{\bf Ribbon operators:} On the Hilbert space ${\cal H}$ one can define so--called ribbon operators \cite{Kitaev,LevinWen,DG16}, which change the graph state in the region covered by the ribbon. From a lattice gauge theory or loop quantum gravity perspective these ribbon operators unify flux and holonomy operators \cite{Kitaev,ABC16a}.  Open ribbon operators require the presence of punctures, we will therefore consider only closed ribbons. For a modular category such as ${\rm SU}(2)_{\rm k}$ the closed ribbon operators are labeled by two representations $(j_o,j_u)$  and  act by inserting  an over--crossing  Wilson loop with colour $j_o$ and an under--crossing Wilson loop with colour $j_u$ along the ribbon \cite{DG16}. 

The type of bases described above diagonalizes the Wilson loops which are parallel to the vacuum loops.  This can be seen by sliding the Wilson loop across the vacuum loop:
\ba\label{DiagWoperators}
\pic{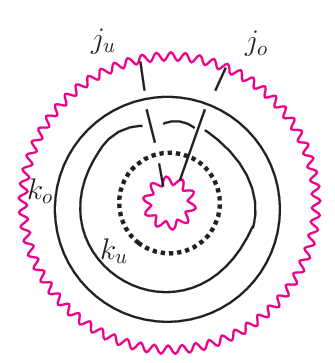}\,=\,\pic{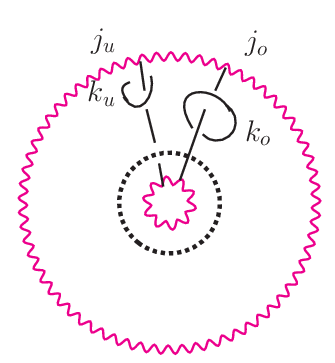}\,=\, \frac{s_{j_ok_o}}{v_{j_o}^2} \frac{s_{j_uk_u}}{v_{j_u}^2} \pic{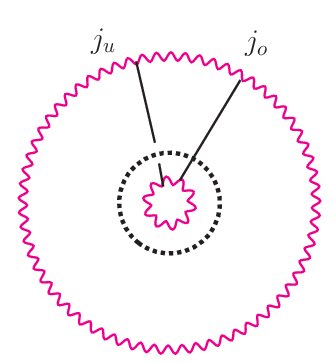}
\ea
where the (rescaled) $S$--matrix $s_{jk}= {\cal D} S_{jk}$ determines the eigenvalues of the Wilson loop operators.

\section{Heegard splittings and diagrams}\label{Sec:Heegard}

Let ${\cal M}$ be a three--dimensional closed, orientable, connected and compact manifold ${\cal M}$. The topology of such a manifold can be encoded into a Heegard diagram \cite{HeegardNotes,GSbook}. Such a diagram is defined as a set of non--contractible and non--intersecting curves on a closed (so--called Heegard) surface $\Sigma$. 

This Heegard surface arises through a Heegard splitting of the manifold ${\cal M}$, that is a representation of ${\cal M}={\cal M}_1 \cup {\cal M}_2$ as the union of two handle bodies ${\cal M}_1$ and ${\cal M}_2$. Handlebodies are three--dimensional manifolds with boundary that arise from the gluing of closed three--dimensional balls. This gluing is accomplished by identifying pairwise disks on the boundary of the 3--balls. 

The Heegard surface $\Sigma=\partial {\cal M}_1 =\partial {\cal M}_2$ is defined as the boundary of these handlebodies and can threfore be considered to be a surface embedded into ${\cal M}$, that splits ${\cal M}$ into two parts. 

The Heegard surface can be equipped with two sets of closed, non--contractible (on $\Sigma$), curves. The first set ${\cal C}_1$ are curves that can be contracted by homotopy in ${\cal M}_1$ to trivial cycles. Here we need only to consider  a minimal generating set of equivalence classes of curves, where two curves are equivalent if they are related by homotopy on $\Sigma$. Likewise, the second set ${\cal C}_2$ is given by (equivalence classes of) curves which can be contracted in ${\cal M}_2$. 

Here we will be interested in representing states on a manifold with defects. These defects will prevent the contractibility of the curves from the set ${\cal C}_1$, that is the defect structure can be identified with the handle body ${\cal M}_1$.

We can achieve a Heegard splitting through a triangulation $\Delta$ of the manifold ${\cal M}$. (Or more generally by using a discretization of ${\cal M}$ via a cell complex.) To this end we consider the one--skeleton $\Delta_1$ of the triangulation, that is the set of edges and vertices. The handlebody ${\cal M}_1$ will be given by the closure of a regular neighbourhood of this one--skeleton. This can be imagined as a blow--up of the one--skleleton.  Indeed, we can construct such a blow--up of a one--skeleton as a handlebody by identifying the blowed up vertices as 3--balls that are glued to each other via the (blowed up) edges. This one--skeleton will be allowed to carry the (curvature) defects by defining curves from the set ${\cal C}_1$ to be not contractible. A (possibly over-complete) set ${\cal C}_1$ of such curves is given by choosing a cycle around each edge of the triangulation.  The (over--) completeness of this set follows from the same argument as we present for the set ${\cal C}_2$ below.

We will  however impose contractibility for the second set of curves ${\cal C}_2$. A  set of curves, which are contractible through ${\cal M}_2$, can be also constructed from the triangulation: We consider the set of triangles in the triangulation and for each triangle $t$ take the curve that arises from $t \cap \Sigma(\Delta)$, where $\Sigma(\Delta)$ is the embedded Heegard surface defined by the triangulation $\Delta$.  This set ${\cal C}_2(\{t\})$ determined by the triangles is (over--) complete: cutting the handlebody ${\cal M}_2$ along disks that are bounded by these curves we remain with a collection of 3--balls, one for each tetrahedron of the triangulation.   The boundary of a given 3--ball  is given by a sphere with four punctures, corresponding to the four triangles bounding the associated tetrahedron. Curves on the sphere not surrounding a puncture can be trivially contracted on this sphere. If a curve surrounds a puncture we can also move this curve across the puncture, as this puncture corresponds to a triangle, and thus a curve in ${\cal C}_2$. 

The Heegard diagram, that is the Heegard surface equipped with the sets ${\cal C}_1$ and ${\cal C}_2$ of curves, encodes the topology of the manifold ${\cal M}$. Such an encoding can be also used to construct a so--called handle decomposition of ${\cal M}$ \cite{GSbook}.  This describes ${\cal M}$ as the gluing of $n$--handles, where $n=0,1,2,3$.  A $0$--handle is a 3--ball and $1$--handles are (`long and thin') cylinders $D \times [0,1]$, that are glued along the discs $D \times \{0\}$ and $D \times \{1\}$ to disks on the boundary of the $0$--handles. The gluing of $0$--handles and $1$--handles gives by definition a handlebody, which will give the first part ${\cal M}_1$ of our Heegard splitting. One can always find a handle--decomposition with only one $0$--handle: if the Heegard surface is a genus $g$ surface we need one $0$--handle and $g$ 1--handles. (In the triangulation picture this reduction can be achieved by contracting the edges of the triangulation along a connected and spanning tree in the graph given by the one--skeleton of the triangulation.) 2--handles are also cylinders (or pancakes) $[0,\varepsilon] \times D$, that are glued along the circumference of the cylinder $[0,\varepsilon] \times \{r=1\}\times [0,2\pi]$ to the handlebody resulting from the $0$-- and $1$--handles.  Given a Heegard diagram one glues the 2--handles along the curves in  the set ${\cal C}_2$ to the handlebody ${\cal M}_1$. (This requires framed curves, but the 2--handles can be only glued along curves without twists, thus a framing is naturally defined parallel to the Heegard surface.) Having glued $0$--,$1$-- and $2$--handles, there is a unique way to glue $3$--handles (solid 3--balls) along their spherical surfaces such that one obtains a closed three--dimensional manifold.

\section{Hilbert space for three--dimensional manifolds with (line) defects}\label{Sec:3+1}

The Heegard diagrams allow us to encode the topology of a manifold into a surface decorated with curves. Additionally we can allow the handlebody ${\cal M}_1$, which can be defined as the blow--up of the one--skeleton of a triangulation, to carry curvature excitations. That is, we do not impose contractibility for  curves in ${\cal C}_1$, that are contractable through the handlebody ${\cal M}_1$, but not in the Heegard surface $\Sigma$. 

We can thus use the Hilbert space structure defined for a surface in section \ref{section:surfaceH} to define a Hilbert space ${\cal H}(\Sigma)$ associated to the Heegard surface $\Sigma$.   Do to the equivalence relations (\ref{StrandDeformation},\ref{Fmove},\ref{bubble move}) for graph based states this Hilbert space encodes degrees of freedom associated to non--trivial cycles of the Heegard surface. Even so we plan to keep cycles contractible through ${\cal M}_1$ as possibly non--trivial, we still want to impose triviality for cycles in ${\cal C}_2$.  

We therefore impose  2--handle constraints. To this end we remind the reader of the sliding property of vacuum loops, discussed in section \ref{section:surfaceH}. This property does indeed ensure that any strand along a vacuum loop can be reduced to a contractible strand. We therefore define projection operators, one for each equivalence class of curves in ${\cal C}_2$. A projection operator associated to a given curve is defined as inserting an over--crossing\footnote{We could also choose to use under--crossing vacuum loops, but over--crossing ones are more natural  as the constraints imposes contractibility through ${\cal M}_2$ that surrounds ${\cal M}_1$.}  normalized vacuum loop along this curve. 

 We then define a new Hilbert space ${\cal H}(\Sigma,{\cal C}_2)$ as the image of these projectors of the Hilbert space ${\cal H}(\Sigma)$. We can also lift operators on ${\cal H}(\Sigma)$, that leave the subspace ${\cal H}(\Sigma,{\cal C}_2)$ invariant, to operators on ${\cal H}(\Sigma,{\cal C}_2)$. 

\subsection{Imposing the 2--handle constraints} \label{sec:2handle}

In section \ref{section:surfaceH} we discussed different bases for the Hilbert space ${\cal H}(\Sigma)$. We now want to choose a basis that simplifies the imposition of the 2--handle constraints.  

A basis for a surface $\Sigma$ can be specified by choosing a set of curves ${\cal C}_B$, along which $\Sigma$ is cut into three--punctured spheres.  Let us choose 
${\cal C}_B$ such that it includes the set ${\cal C}_2$.
As explained in section \ref{Sec:Heegard} these curves cut $\Sigma(\Delta)$ into four--punctured spheres. (If we use other lattices we may obtain spheres with a different number of punctures.) Thus all that remains to do is to choose for each of these spheres one closed curve that cuts a given four--punctured sphere into two three--punctured ones. Note that these additional cycles can be generated from the cycles in ${\cal C}_2$. 

The basis associated to the set ${\cal C}_B$ is then constructed by $(a)$ inserting a vacuum loop along each loop in ${\cal C}_B$. (It is actually sufficient to consider only vacuum loops along curves in ${\cal C}_2$, as the corresponding cycles generate all the cycles in ${\cal C}_B$.) And $(b)$ we construct a graph ${\cal F}$ dual to the curves in ${\cal C}_B$. We then consider a doubling of this graph ${\cal F}$; one copy ${\cal F}_u$ made of under-crossing strands and the other copy ${\cal F}_u$ of over-crossing strands.  In particular ${\cal F}_u$ under-crosses the vacuum lines along curves in ${\cal C}_B$ and ${\cal F}_o$ over-crosses the same vacuum lines. Allowing all admissible colourings of these graphs we obtain a basis for ${\cal H}(\Sigma)$. 

We now impose the 2--handle constraints, that is impose the projectors given by over-crossing vacuum loops along curves in ${\cal C}_2$. Along a given curve $c$ in ${\cal C}_2$ we thus have two vacuum loops: one, $c_o$ is over-crossing the dual strand in both graphs ${\cal F}_u$ and ${\cal F}_o$, and the other vacuum loop $c_u$ is over-crossing ${\cal F}_u$ but under-crossing ${\cal F}_o$.  We can slide the vacuum loop $c_u$ over $c_o$. This does not affect ${\cal F}_u$, as it under--crosses both vacuum loops.  However after the sliding, the vacuum loop $c_u$ is encircling the strand  of ${\cal F}_o$ which is dual to $c$. Hence according to the killing property (\ref{annihil}) the strand is annihilated, i.e. only states in which this strand of ${\cal F}_o$ carries the trivial label, survive:
\ba\label{Impose2H}
\pic{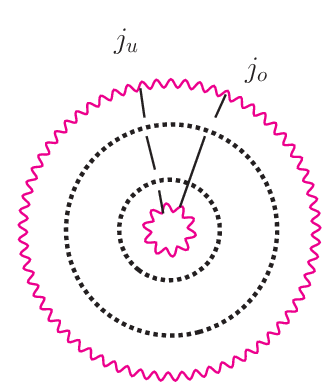}\,\,=\,\,\pic{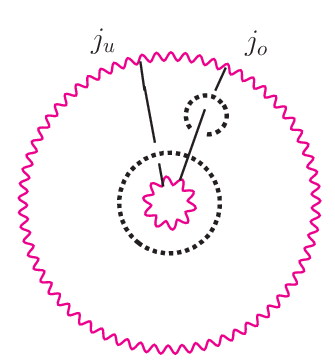}\,\,=\,\, {\cal D}\, \, \delta_{j_o0}\,\,\pic{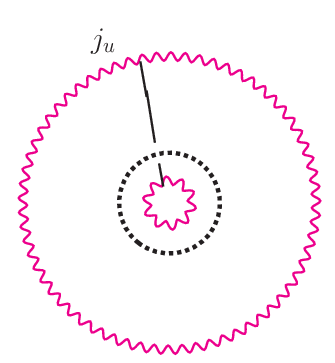} \q .
\ea

This annihilation applies to all strands of the over--crossing copy ${\cal F}_o$, that cross loops in ${\cal C}_2$. The cycles in ${\cal C}_2$ do generate however also the remaining cycles in ${\cal C}_B$, and in fact the entire copy ${\cal F}_o$ is annihilated. That is all strands of this copy ${\cal F}_o$ are forced to carry the trivial representation label $j=0$. 

Thus we have a basis of the constrained Hilbert space ${\cal H}(\Sigma, {\cal C}_2)$ labelled by all admissible colourings of the graph ${\cal F}_u$, which is dual to a set of cutting curves ${\cal C}_B$ obtained from completing the set ${\cal C}_2$.  We will refer to such a basis as ${\cal B}_2$--basis.   

We have seen that the 2--handle constraints kill the over--crossing copy of the graph ${\cal F}_o$. The resulting state space can be identified with the state space for the WTR model. Indeed, the TV partition function is in a precise sense given by a square of the WTR partition function \cite{BalKir,Virelizier}.

\subsection{Bases and Fourier transform}\label{Sec:Ftrafo}

We can express this set of states, satisfying the 2--handle constraints, also in an alternative basis. As explained in section \ref{section:surfaceH} a basis transformation can be implemented via ${\mathbb S}$ and ${\mathbb F}$ maps.  Both these maps do not mix the under-crossing and over-crossing copies ${\cal F}_u$ and ${\cal F}_o$. Let us consider an ${\mathbb S}$--move, defined in equ. (\ref{SStrafo}), and apply it to some element of the ${\cal B}_2$ basis, satisfying the 2--handle constraints:
\ba\label{StrafoC}
\pic{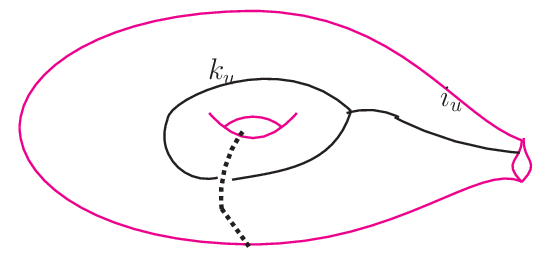} \,\,=\,\,\sum_{j_u} B^{i_u}_{k_uj_u} \pic{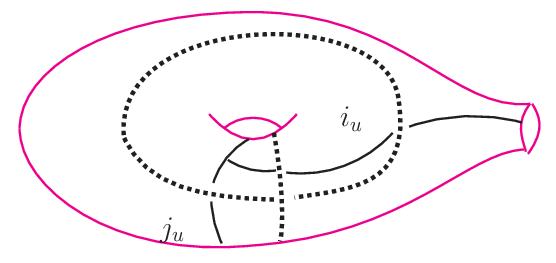} \q .
\ea
Here we assume on the left hand side a state of the ${\cal B}_2$ basis, with a 2--handle constraint along the meridian curve of the punctured torus.  Thus the over--crossing graph copy is labelled by trivial representations, in particular $k_o=0$. Exchanging the meridian curve with the equator curve in the set ${\cal C}_B$, which defines a new basis, we can express the  state of the original basis, shown on the left hand side of (\ref{StrafoC}) as a linear combination of new basis states, shown on the right hand side of  (\ref{StrafoC}). Here the over--crossing graph copy ${\cal F}_o$ features a vacuum loop along the curve that defines the 2--handle constraint.

We see that the ${\mathbb S}$--move does define new basis states that satisfy again the 2--handle constraints. These basis states are again labelled by the representations assigned to the under--crossing graph copy only. The over--crossing graph features a vacuum loop along the 2--handle attaching curve. 

Follow--up ${\mathbb F}$--moves can in principle lead to a more involved prescription for the over--crossing graph copy ${\cal F}_u$. We can however always re--apply the (normalized) vacuum loops along the ${\cal C}_2$ curves, and the sliding property for these vacuum loops can be used to simplify the over--crossing graph copy again. In addition one can use the sliding property across the vacuum loops which define the basis, i.e. are parallel to the ${\cal C}_B$ curves.  
We will however not need to know the exact state for the over--crossing copy: it suffices to know that it is uniquely prescribed (as it is defined from applying the basis transformation to a ${\cal B}_2$ basis state where the over--crossing copy ${\cal F}_u$ is trivial), and that the 2--handle constraints are always imposed.

Above we discussed a basis constructed from cutting curves that include the set ${\cal C}_2$ of 2--handle attaching curves. These curves go along the boundaries of the triangles. The graph ${\cal F}$ is dual to these curves: thus we have one link for each triangle. These links meet in a priori four--valent nodes, representing the tetrahedra of the triangulation. For the graph ${\cal F}$ these nodes are expanded into three--valent nodes. Thus the ${\cal B}_2$ basis can be identified as a quantum deformed spin network basis \cite{RovelliSmolin, SmolinQ,lqg}. This allows us to define a (Ashtekar--Lewandowski like) vacuum state (on a fixed triangulation), by assigning only trivial labels also for the under--crossing graph ${\cal F}_u$ in the ${\cal B}_2$ basis. 

Another interesting basis is constructed from cutting curves that include the set ${\cal C}_1$, given by cycles around all the edges of the triangulation. If we cut the Heegard surface along all curves in ${\cal C}_1$ we remain with punctured spheres associated to the blown--up vertices of the triangulation.  Depending on the valency of these vertices we need to introduce further cutting curves to obtain three--punctured spheres. These additional cutting curves are generated by the cycles in ${\cal C}_1$ however.  We refer to this basis as ${\cal B}_1$ basis.

Note that, for the case that the three--dimensional manifold ${\cal M}$ has the topology of the 3--sphere,  the 2--handle constraints together with the vacuum loops parallel to the ${\cal  C}_1$ cycles allow to reduce the over--crossing graph copy to the vacuum loops associated to the 2--handles.

We can use the ${\cal B}_1$ basis, determined by ${\cal C}_1$, to impose flatness constraints, i.e. make also curves in ${\cal C}_1$ contractible, and thus remove defect excitations a-priori allowed on ${\cal M}_1$. These flatness constraints define the plaquette terms in the Walker--Wang Hamiltonian \cite{WW}.  In fact, as we will discuss in the next section Wilson loops along ${\cal C}_1$ curves will be diagonalized by the ${\cal B}_1$ basis.  In this sense the ${\cal B}_1$ basis diagonalizes the Walker--Wang Hamiltonian\footnote{Here we consider only the plaquette part of the Hamiltonian. A second part suppresses violations of the coupling conditions at the nodes, but we do not consider such violations here.}  and labels the (curvature) excitations of the Walker--Wang model. 

Ground states of the Wallker--Wang model have to satisfy the flatness constraints. These are imposed by under-crossing vacuum loops around the edges of the triangulation, that is along the curves in ${\cal C}_1$. We thus have for each edge of the triangulation two parallel vacuum loops, one under-crossing  the graph ${\cal F}_u$ and the other over-crossing  the graph ${\cal F}_u$. These vacuum loops force all representation labels of ${\cal F}_u$ to be trivial.

Hence we have shown that the ground state of the model as defined here, is unique, namely given by a labelling of ${\cal F}_u$ with only trivial representations.  (This holds for modular fusion categories where the killing property (\ref{annihil}) holds.) The line of arguments we used here is quite straightforward compared to the arguments used in \cite{Burnell}, to show the non--degeneracy of the vacuum of the 3--torus.\footnote{A similar argument is used in \cite{Burnell2} to show uniqueness of the vacuum for a specific model based on a cubical lattice.}  Additionally our argument covers all compact topologies (of the equal--time hypersurfaces) at once. 

We conjecture that this ground state, i.e. the state given by assigning trivial labels to ${\cal F}_u$ in the ${\cal B}_1$ basis, can be also identified with a quantum deformed BF or Crane--Yetter vacuum.

We thus have two dual vacuum states, that are both given by trivial representation labels for the under--crossing graph copy.  These two vacuum states are associated to graphs that are given by the one--skeleton of the triangulation (for the quantum deformend BF vacuum) and the one--skleleton of the dual complex (for the quantum deformed AL vacuum) respectively.

\subsection{Operators on the constrained Hilbert space} \label{operatorsC}

We can furthermore consider operators on ${\cal H}(\Sigma)$ that commute with the projectors defined by the 2--handle constraints, that is operators that leave the constrained Hilbert space invariant.  In section \ref{section:surfaceH} we discussed (closed) ribbon operators, which in the case of the modular fusion category ${\rm SU}(2)_{\rm k}$ amount to inserting over-crossing and under-crossing Wilson loops. From the discussion in section \ref{sec:2handle} it is clear that the under-crossing Wilson loops preserve the constrained Hilbert space ${\cal H}(\Sigma, {\cal C}_2)$. 

There are two particularly interesting classes of loops: along curves in ${\cal C}_1$ and along curves in ${\cal C}_2$. Let us discuss the loop operators along these curves, starting with curves from ${\cal C}_1$. As shown in (\ref{DiagWoperators})  loop operators along ${\cal C}_1$ curves are diagonalized by a ${\cal B}_1$ basis, for which we have vacuum loops along the ${\cal C}_1$ curves.  Similarly Wilson loops along ${\cal C}_2$ curves will be diagonalized by the ${\cal B}_2$ basis. 

As discussed the ${\cal B}_2$ basis corresponds to the spin network basis as e.g. used in loop quantum gravity. A $\SU(2)$ spin network is a graph based on the dual graph to a triangulation, which is labelled by representations of $\SU(2)$.  The spin network  basis diagonalizes the Casimir operator formed from the so--called flux operators, associated to a given triangle. Geometrically, the square root of this Casimir operator defines the area operator for this triangle, and has eigenvalues proportional to $\sqrt{j(j+1)}$, if the link through this triangle carries the label $j$ \cite{RovelliSmolin}.  In the quantum group case the $k$--Wilson loop operator along a ${\cal C}_2$ curve that (in the ${\cal B}_2$ basis) crosses a strand labelled by $j$, gives an eigenvalue
\ba
\frac{s_{jk}}{v_j^2}&=&\frac{(-1)^{2k}[(2j+1)(2k+1)]}{ [2j+1]} \,=\, (-1)^{2k} \frac{\sin\left(  \frac{\pi}{{\rm k}+2} (2j+1)(2k+1) \right)}{\sin\left(   \frac{\pi}{{\rm k}+2} (2j+1)    )\right)}  \q .
\ea
If we consider a normalized $k$--Wilson loop operator, that is a $k$--Wilson loop operator divided by the (signed) quantum dimension $v_k^2$ we obtain as eigenvalue
\ba\label{limitflux}
\frac{\sin\left(  \frac{\pi}{{\rm k}+2} (2j+1)(2k+1) \right)  \sin\left(  \frac{\pi}{{\rm k}+2}\right) }{\sin\left(   \frac{\pi}{{\rm k}+2} (2k+1)    )\right) \sin\left(   \frac{\pi}{{\rm k}+2} (2j+1)    )\right)} \q \stackrel{{\rm k} \rightarrow \infty} \longrightarrow \q1 - \tfrac{8}{3} \, j(j+1) \, k(k+1) \, \left( \tfrac{\pi}{{\rm k}+2} \right)^2 \q .
\ea
Thus we can extract in the limit the $\SU(2)$ Casimir eigenvalue $j(j+1)$. in the loop quantum gravity interpretation \cite{RovelliSmolin} this  gives the square of the area of the triangle  enclosed by the Wilson loop. The limit (\ref{limitflux}) for the eigenvalues of the normalized Wilson loop operators suggest a geometric interpretation for these operators: In $\SU(2)$ one can approximate the Casimir operator of the group via a sum over the cosine function applied to the  Lie algebra generators \cite{DG}. Due to the exponentiation this operator does however violate gauge invariance. One can however project the operator back to a gauge invariant one, and the resulting spectrum approximates the Casimir spectrum $j(j+1)$ for sufficiently small representation labels \cite{DG}. For larger $j$ the bound imposed by taking the cosine of the generators sets in. 

The eigenvalues (\ref{limitflux}) show also this behaviour, with the representation $k$ of the Wilson loop functioning as the exponentiation parameter and playing the same role as a step size for a discrete Laplacian. Thus we can relate the normalized Wilson loop around the triangles to an (exponentiated) area operator.  Here the gauge invariance is manifestly preserved however.

Another reason to identify the Wilson loop around triangles with an exponentiated area operator is the analysis of \cite{HHKR}. There one considers phase spaces associated to homogeneously curved simplices.  The holonomy around a homogeneously curved triangle is thus constraint by the fact that the curvature integrated over the triangle has to be proportional to the area of the triangle. A third reason is provided by \cite{3to4}, which constructs the lift of state spaces and operators for three--dimensional TQFT's to state spaces and operators for four--dimensional TQFT's with line defects to BF theory with classical groups. In this case ribbon operators that go around triangles and preserve the 2--handle constraints map indeed to (exponentiated) flux operators, from which one can define via gauge averaging the (exponentiated) area operators.

For (normalized) Wilson loops along ${\cal C}_1$ curves, that is around the edges of a triangulation,  one finds -- of course -- the same eigenvalues. These operators are diagonalized in the ${\cal B}_1$ basis \cite{DG16}.  These eigenvalues (\ref{limitflux}) do indeed approach in the limit ${\rm k} \rightarrow \infty$ the eigenvalue for the normalized $\SU(2)$ Wilson loop operator (with representation label $k$)
\ba
\frac{\sin( (2k+1) \theta)}{(2k+1) \sin(\theta)}
\ea
for a state peaked on a curvature (class) angle $\theta$ along the encircled edge.  Thus curvature is discretized as we can identify $\theta=\tfrac{\pi}{{\rm k}+2}(2j+1)$. 

The fact that (exponentiated) area operators and curvature operators have the same eigenvalues hints towards a  duality relation, see also \cite{BarrettObs}. We conjecture that this fact is due to  the polar duality \cite{kokkendorff} for spherical simplices:  For a given spherical simplex $\sigma$ one can construct a dual simplex $\sigma'$ whose lenghts are determined by the dihedral angles of $\sigma$.

\section{Examples}\label{examples}

Here we will consider a number of examples of triangulations and topologies. 

\subsection{A defect along a loop}

We start with a simple example and choose ${\cal M}$ to be the 3--sphere with a  defect loop inserted. We therefore define a genus 1 Heegard splitting of the sphere: The handle--body ${\cal M}_1$ is given by a solid torus, its equatorial line defines the loop carrying the defect. (Thus loops encircling this defect line are not contractible.) The Heegard diagram includes also an attaching curve (below as red dashed line) for the 2--handle which is along the equator of the torus:
\ba
\pic{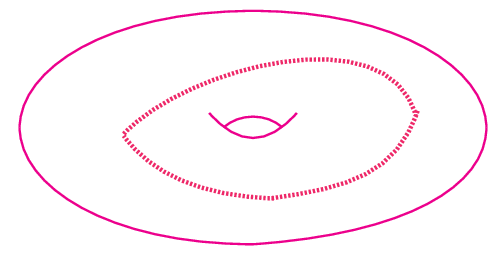}
\ea
The ${\cal B}_1$ and ${\cal B}_2$ basis, after the 2--handle constraints have been imposed, are as follows:
\ba
{\cal B}_1(i)\,:=\, \pic{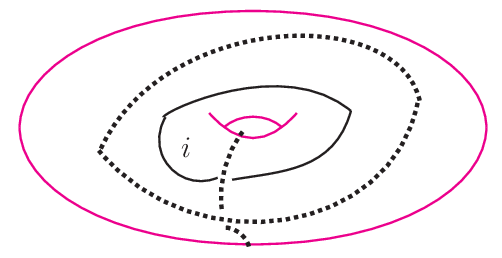} \q,\q     {\cal B}_2(j)\,:=\, \pic{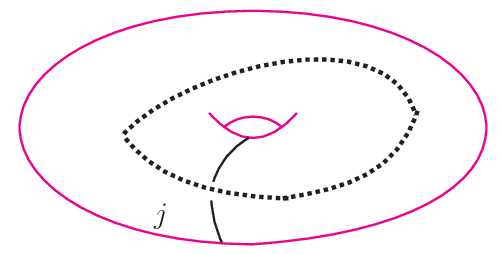} . \q
\ea
The two bases are connected by an $S$--transformation
\ba\label{Sthurs}
{\cal B}_1(i) = \sum_j S_{ij} \, {\cal B}_2(j) \q . 
\ea
We consider  two kinds of operators: under-crossing Wilson loops along meridians and equators of the torus. The meridian $k$--Wilson loops $W_{\rm meri}(k)$ are diagonalized by the ${\cal B}_1$ basis, whereas they generate the ${\cal B}_2$ basis from the state ${\cal B}_2(0)$:
\ba
W_{\rm meri}(k) \rhd \, {\cal B}(0) &=&   {\cal  B}(k) 
\ea
Analogously the equatorial Wilson loops are diagonalized by the ${\cal B}_2$ basis, whereas they generate from the vacuum ${\cal B}_1(0)$ the states of the ${\cal B}_1$ basis.

Note that a ${\cal B}_1(0)$ basis state is expressed via the $S$--transformation as the following combination of ${\cal B}_2$ states:
\ba
\sum_j  \frac{v_j^2}{\cal D} {\cal B}_2(j) \q ,
\ea
which describes a vacuum loop along the meridian under--crossing a vacuum loop along the equator.  The ${\cal B}_2(0)$ state is equivalent to the state given by a pair of vacuum loops along the equator over and under--crossing a vacuum loop along the meridian.  Indeed using the sliding and annihilation property of the vacuum loops one can show that such a state reduced to the state shown on the right hand side of   (\ref{Sthurs}). 

\subsection{The two--tetrahedra triangulation of the 3--sphere}

\begin{figure}[!]
	\centering
	\begin{minipage}[b]{1\textwidth}
		\centering
		\includegraphics[scale = 0.5]{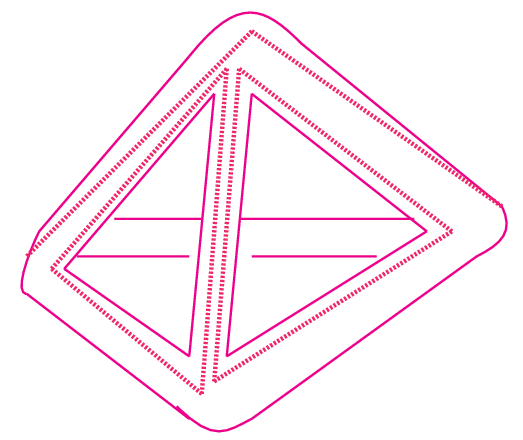}	\q \q\q  \includegraphics[scale = 0.5]{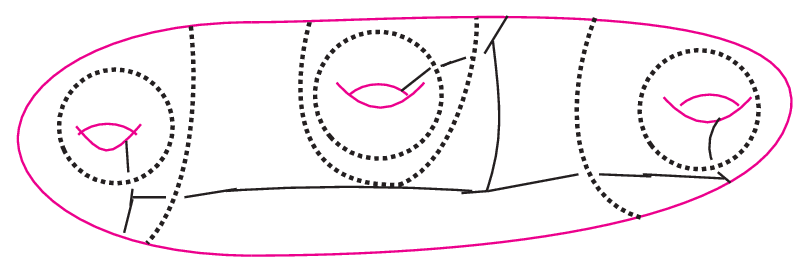}	
	\end{minipage}
	\caption{The left panel shows a Heegard diagram for the 3--sphere, determined by a triangulation consisting of two tetrahedra. Three attaching curves, corresponding to three of the triangles are depicted as red dashed lines.  The attaching curve associated to the fourth (bottom) triangle can be obtained from  combining the other three curves. \\
	On the right panel the Heegard surface is deformed into the standard form for a genus 3 surface. Three 2--handle constraints are imposed via vacuum loops. The 2--handle constraint for the fourth triangle (determined by a curve surrounding all three holes) follows from the other three constraints. To see this one has to use the sliding property of the vacuum loops. We have also indicated  a basis, which is neither a ${\cal B}_1$ basis nor a ${\cal B}_2$ basis.
	 \label{fig:tetra1}}
\end{figure}

We again consider the 3--sphere, this time triangulated by two tetrahedra, that are glued to each other. The gluing identifies the vertices, edges, and triangles of the two tetrahedra with each other. The one--skeleton of this triangulation coincides therefore with the one--skeleton of one tetrahedron. This gives a genus 3 Heegard surface, see figure \ref{fig:tetra1}. The three loops around the holes can be identified with three independent 2--handle attaching curves, corresponding to three of the four triangles. (The 2--handle curve associated to the fourth triangle is generated by the other three curves.) 

\begin{figure}[!]
	\centering
	\begin{minipage}[b]{1\textwidth}
		\centering
		\includegraphics[scale = 0.5]{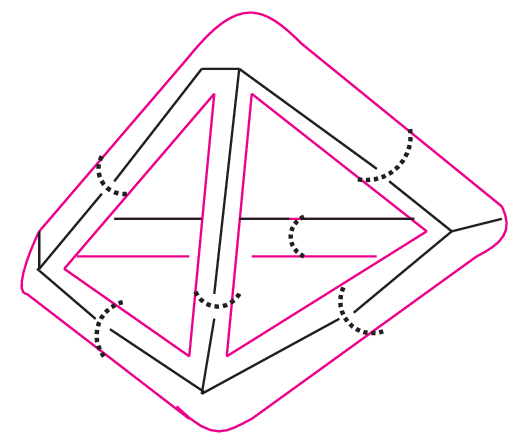}	\q\q \q  \includegraphics[scale = 0.5]{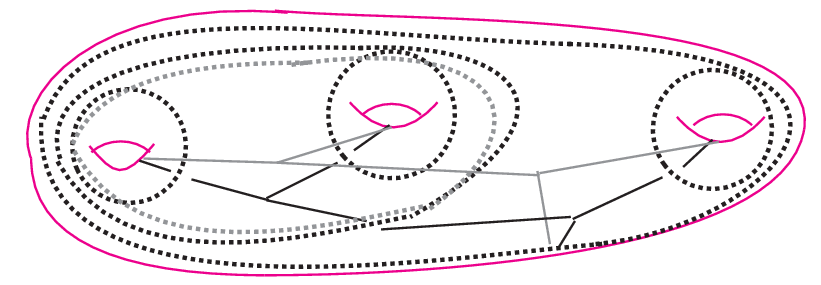}	
	\end{minipage}
	\caption{The left panel shows a ${\cal B}_1$ basis for the two--tetrahedra--triangulation of the 3--sphere.  Here we did not depict the over--crossing graph copy ${\cal F}_u$ and the vacuum loops associated to the 2--handle constraints. This copy can be transformed, using sliding across the 2--handle vacuum loops, to the same 2--handle vacuum loops. The basis is labelled by six quantum numbers and diagonalizes Wilson loop operators around the (six) edges of the triangulation.
	 \\
	The right panel shows a ${\cal B}_2$ basis. The strands and vacuum loop on the backside of the genus 3 surface are depicted in grey. Note that for both bases we have 6 vacuum loops. But always three vacuum loops can be generated from the other three loops, using the projection property (modulo factors of ${\cal D}$) and the sliding property of the loops.  The graph ${\cal F}_u$ can be identified with the one--skeleton of the dual complex to the triangulation: two  four--valent nodes (expanded into three--valent ones) representing the two tetrahedra and connected with each other by four links, representing the four triangles. This basis diagonalizes Wilson loop operators around the four triangles of the triangulation and in addition two Wilson loops around pairs of triangles.
	 \label{fig:tetra2}}
\end{figure}

We can construct  a ${\cal B}_1$ basis and a ${\cal B}_2$ basis.   The ${\cal B}_2$ basis corresponds to (an expansion to three--valent nodes of) the usual spin network basis, which is based on graphs with nodes dual to tetrahedra and links dual to triangles.  These two bases are shown in figure \ref{fig:tetra2}.  

Following  the discussion in section \ref{operatorsC} we can interpret the ${\cal B}_1$ basis as diagonalizing the curvature around the edges of the tetrahedron, and the ${\cal B}_2$ basis as encoding the spatial geometry, i.e. the four areas and one dihedral angle per tetrahedron. 

A similar relation between the triangulation and the dual complex for a tetrahedron was noted in \cite{BarrettObs}, which discusses geometric expectation values for the Turaev-Viro partition function. This lead to a duality (or Fourier) transformation for the quantum $\{6j\}$ symbols, where the transformation is defined via the $S$--matrix. Here we identified a similar Fourier transform as basis transformations in a Hilbert space describing a $(3+1)$--dimensional theory.

\subsection{The 4--simplex triangulation of the 3--sphere}

\begin{figure}[!]
	\centering
	\begin{minipage}[b]{1\textwidth}
		\centering
		\includegraphics[scale = 0.5]{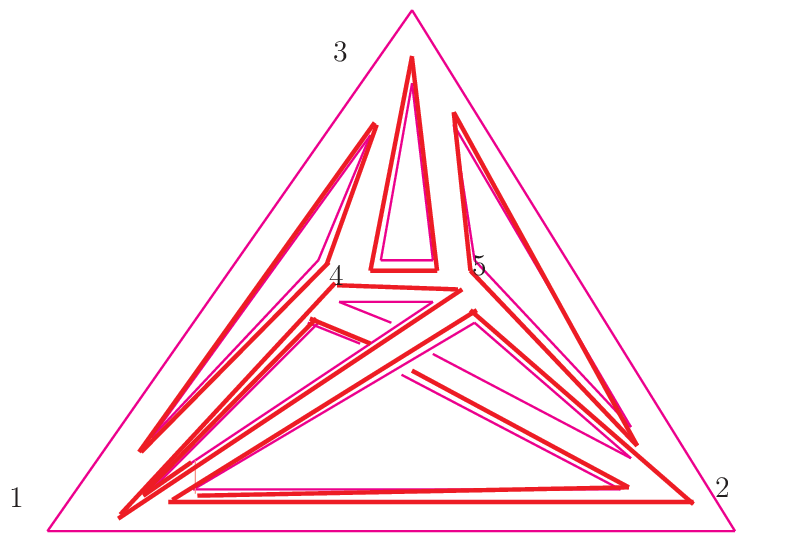}	\q\q \q  \includegraphics[scale = 0.5]{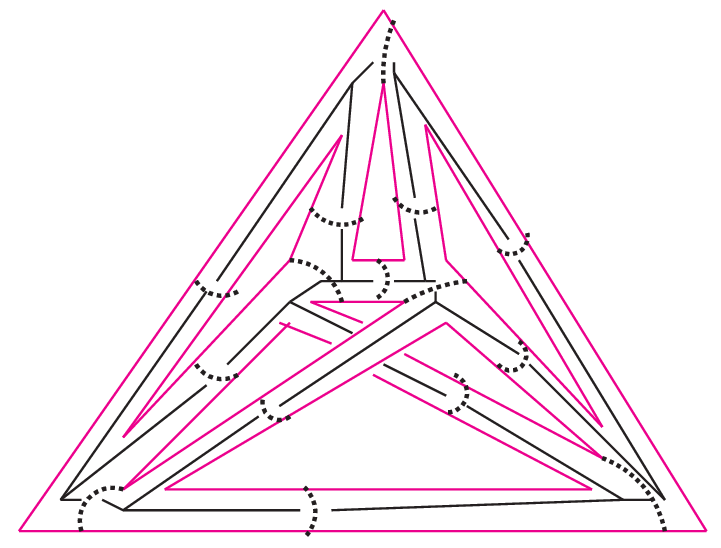}	
	\end{minipage}
	\caption{The left panel shows the Heegard diagram for the 4--simplex triangulation. The Heegard surface is of genus six. Six of the ten attaching curves for the 2--handles are shown. The other four curves are generated from these six curves.	Note that the curve for the triangle $t(124)$  under--crosses the handle representing the edge $e(15)$.\\
	The right panel shows a ${\cal B}_1$ basis for the 4--simplex triangulation. For each vertex of the triangulation one can choose a cutting of the associated four--punctured sphere into two three--punctured ones. Again we have not shown the over---crossing graph copy. Using the 2--handle vacuum loops, as well as the vacuum loop around the edge $e(15)$, one can transform the over--crossing graph copy to the 2--handle vacuum loops.
	 \label{fig:4simplex}}
\end{figure}

The next simplest triangulation of the 3--sphere is given by the boundary of a 4--simplex. The resulting Heegard surface is of genus 6 and depicted in figure \ref{fig:4simplex}. The 4--simplex triangulation is self--dual, that is the dual complex has the same connectivity as the 4--simplex.   A ${\cal B}_1$ basis is depicted in figure \ref{fig:4simplex}.  

The ${\cal B}_1$ basis diagonalizes Wilson loop operators around the ten edges and also Wilson loop operators around certain pairs of edges.  The ${\cal B}_2$ basis diagonalizes Wilson loop operators around the ten triangles and around certain pairs of triangles.  

Such a dual structure of holonomy operators, around both edges and triangles of a 4--simplex, plays also an important role in \cite{HHKR}, which constructs the 4--simplex amplitude for a spin foam model describing gravity with a cosmological constant, and examines the semi--classical limit of this amplitude. \cite{HHKR} also discusses the phase space associated to the boundary of a homogeneously curved simplex, and notes the dual role of the two kinds of holonomies. We conjecture that the Hilbert spaces and operators constructed in this work provide a quantization of the phase spaces discussed in \cite{HHKR}.  

\subsection{The 3--torus}

\begin{figure}[!]
	\centering
	\begin{minipage}[b]{1\textwidth}
		\centering
		\includegraphics[scale = 0.5]{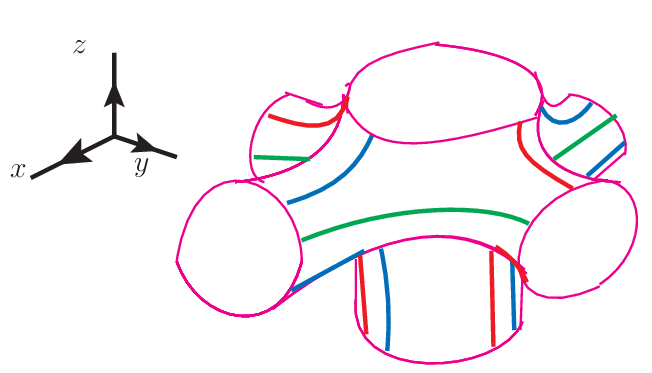}	\q\q \q  \includegraphics[scale = 0.5]{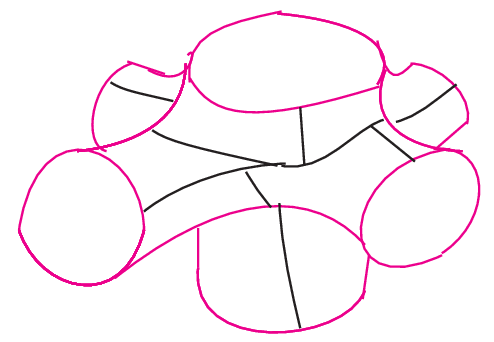}	
	\end{minipage}
	\caption{The left panel shows a Heegard diagram for the 3--torus.  The $x,y$ and $z$ directions have to be periodically identified. The  Heegard surface is based on a lattice constructed from a cube whose sides have been identified pairwise. This surface is of genus three: there is a punctured sphere corresponding to one vertex of the triangulation and three one--handles corresponding to the three edges of the lattice. There are three two--handles corresponding to the three pairs of identified sides. The attaching curves  are shown in blue, green and red. To obtain the Heegard surface for a lattice with more cubes we just need to glue more of the basic building blocks, shown in the panel, to each other. \\
	The right panel shows a  choice for a ${\cal B}_1$ basis, where for clarity we have omitted  vacuum loops and the over--crossing graph ${\cal F}_u$.	 \label{fig:3torus}}
\end{figure}

Next we consider the 3--torus, but this time we work with a cubical lattice. To begin with we choose the smallest possible lattice consisting of one cube, three edges and one vertex. The dual complex is given by the same cubical lattice. The Heegard diagram for such a lattice is depicted in figure \ref{fig:3torus}, as well as a choice for the ${\cal B}_1$ basis.

The ${\cal B}_1$ basis diagonalizes Wilson loops around the edges of the (direct) lattice -- or alternatively around the plaquettes of the dual lattice, whereas the ${\cal B}_2$ lattice diagonalizes loops around the faces of the direct lattice, or around the links of the dual lattice. 

The  basis allows us to identify the 3--torus Hilbert space (also with a more refined lattice) with the one proposed by Walker and Wang \cite{WW}. The Walker--Wang models generalize the Levin--Wen string net models \cite{LevinWen} from $(2+1)$ dimensions to $(3+1)$ dimensions. The Hilbert space is defined as a span of states based on a cubical lattice, whose nodes have been expanded to three--valent nodes. Then plaquette operators are introduced by defining Wilson loop operators for the plaquettes of the lattice together with their planar projections, so that the graphical equivalences (\ref{Fmove},\ref{bubble move},\ref{resolution of crossing}) can be used. This can be done such that the plaquette operators commute.

It is clear that such a prescription is also provided by the ${\cal B}_2$ basis. (The ${\cal B}_1$ basis would also provide such a prescription, but one traditionally understands under `plaquette' operators Wilson loop operators around the faces of the dual lattice, or equivalently around the edges of the direct lattice. Thus we identify the original prescription of the Walker--Wang model with the ${\cal B}_2$ basis.) The plaquette operators are given by (under--crossing) Wilson loops around the edges of the lattice, which  -- as the loops are not intersecting on the Heegard surface -- clearly commute. In addition we have Wilson loop operators around the links of the dual lattice, which are however diagonal in the ${\cal B}_2$ basis.  An important difference between the version of the Walker--Wang model discussed here and the original definition \cite{WW}, is the presence of over--crossing vacuum loops. In our view this feature leads however to an improved and more consistent version of the model. This feature allows in particular the straightforward identification of the (unique) ground state. 

Our treatment provides also a dual basis for the Walker--Wang model, namely the ${\cal B}_1$ basis. This basis diagonalizes the plaquette operators, which in the definition of the Walker--Wang model, contribute to the Hamiltonian operator. (A second part of the Hamiltonian measures gauge invariance violations at the nodes, which we here do not consider.)  The ${\cal B}_1$ basis therefore diagonalizes the Hamiltonian of this model. 
As we have argued in section \ref{Sec:Ftrafo} the ground state for this Hamiltonian is unique and given by the state of the ${\cal B}_1$ basis where all representation labels of the under--crossing graph copy are set to be trivial. 

This argument generalizes to other topologies and triangulations and is much simpler as the one provided in \cite{Burnell} for the periodic cubical lattice. The uniqueness of the ground state depends crucially on the modularity of the fusion category, as we use the killing property (\ref{annihil}) for the vacuum loops. Indeed non--modular fusion categories (e.g. fusion categories based on group representations) can lead to ground state degeneracy \cite{Burnell}. The (generalized) fusion bases for such non--modular categories will be discussed in \cite{BCtoappear}. Another possibility to obtain ground state degeneracy is to apply the strategy of B\"arenz and Barrett \cite{BB4D} for the construction of new so--called dichromatic invariants for four--dimensional manifolds. The construction of these invariants is  based on a handle decomposition of the four--dimensional manifolds. Also for four--dimensional manifolds such decompositions are determined by the 1--handles and 2--handles only.  For the Crane--Yetter invariant one decorates both the 1--handles and 2--handles with vacuum loops. For the more general dichromatic invariants the  2--handles are not necessarily decorated with the vacuum loop of the full fusion category, but with a vacuum loop of a subcategory. More generally one can invoke a second fusion category and a functor between the two fusion categories.  The same strategy can be applied here: we could replace the constraints, so far given by vacuum loops of a modular fusion category along the attaching curves of the two--handles, by e.g. vacuum loops of a subcategory. Such `dichromatic' models may also lead to ground state degeneracy as shown in \cite{BB4D}.

\section{Discussion}\label{discussion}

In this work we constructed state spaces and operators for $(3+1)$--dimensional topological field theories with line defects. To this end we utilized the encoding of a  triangulated three--dimensional manifold via a Heegard surface which is equipped with a set of  so--called 2--handle attaching curves. This allowed us to start from the state spaces and operators of a three--dimensional topological quantum field theory. To obtain the state space of a $(3+1)$--dimensional theory we have in addition to impose constraints related to the 2--handle attaching curves. This yielded in particular an improved version of the Walker--Wang model \cite{WW}.

Here we considered as $(2+1)$--dimensional starting point the Turaev--Viro topological quantum field theory \cite{TV}, based on a modular fusion category. This construction can be generalized to pre--modular (or spherical) fusion categories, indeed \cite{3to4} considered the representation category of a finite group.    Due to the modification of the killing property (\ref{annihil}) in this case, the structure of excitations is more involved. In particular the associated Walker--Wang Hamiltonian will feature a ground state degeneracy  \cite{ Burnell}.  An analysis of the ground states and more generally the construction of the generalized fusion bases in the case of pre--modular categories will appear elsewhere \cite{BCtoappear}. Another generalization is to follow the strategy of B\"arenz and Barrett \cite{BB4D} in replacing the 2--handle constraints with (less restrictive) constraints. This would presumably lead to state spaces associated to the dichromatic invariants constructed in \cite{BB4D}, but also allow for line defects.  Thus the technique here could allow for a large variety of state spaces, describing $(3+1)$--dimensional TQFT's with defects, see \cite{dominique} for alternative constructions.
The advantage of the method pursued here is that it allowed us to construct an interesting set of bases, resulting from the $(2+1)$--dimensinoal fusion basis. One of these bases diagonalizes the Walker--Wang Hamiltonian. This makes it straightforward to prove ground state uniqueness (for the case of modular fusion categories) and to understand the properties of the excitations of this Hamiltonian.

In a quantum gravity context the construction here develops and makes much more concrete ideas expressed in e.g. \cite{smolin1,smolin2}. In \cite{smolin2} one also uses surfaces to represent $(3+1)$--dimensional state spaces. The surfaces are equipped with a spin network but one does not impose 2--handle constraints.  The work here reemphasizes the role of topological quantum field theory in quantum gravity \cite{smolin1,TimeEvol} and paves the way to realize quantum gravity as a theory of space time foam.  It connects to the recent construction of spin foam simplex amplitudes via the boundary Chern--Simons theory \cite{HHKR}. It provides a quantization for the phase space constructed in \cite{HHKR}.

A particular interesting aspect found in this work are the self--dual features of the quantum geometry states: firstly, in the set of generalized fusion bases there are two bases that are dual to each other and diagonalize quantum deformations of holonomy operators or quantum deformations of area operators respectively. Secondly both operators are implemented via Wilson loops and the spectra of these two classes of operators do agree.  

Apart from the generalization of the models as described above there are many directions of research to follow up:

~\\
{\bf Introduction of boundaries:} In some sense we treated already the Heegard surface as boundary of the handle--body ${\cal M}_2$. This is because we impose that curves can be contracted through ${\cal M}_2$ but not through the complement handle--body ${\cal M}_1$. One question would be to generalize this set--up and  use compression bodies instead of handle--bodies \cite{HeegardNotes}. This allows a Heegard splitting for three--manifolds with boundary.  Another way to introduce boundaries is to allow cuts through the Heegard surface leading to punctures of this surface. One would then start with the state space of e.g. the Turaev--Viro model with punctures \cite{KKR,Kir,DG16}. This would include -- in addition to curvature defects -- torsion defects, that is violations of the Gau\ss~constraints that are implemented by allowing strands to end at the boundary of the Heegard surface. These kind of cuts would also allow the definition of gluings of state spaces and the related construction of entanglement entropy as in \cite{ABC16b}. Such a gluing could be performed in the ${\cal B}_1$ basis (in which triangulation edges would be glued) or in the ${\cal B}_2$ basis  (in which spin network links would be glued), which can lead to different notions of entanglement entropy.  Another treatment of boundaries and (string--like) defects in the context of (undeformed) BF theory can be found in \cite{BaezPerez}. There the string--like defects carry a group--valued field that couples to the spin network describing the BF theory of the bulk.

Having at our disposal different ways of introducing boundaries the next question is to impose boundary conditions and to analyze the associated boundary excitations. E.g. the analyses \cite{Burnell} shows the existence of surface  anyons with a certain choice of boundary conditions.

~\\
{\bf Refinement moves and vacua:} Here we based the Heegard surface on a fixed triangulation. One can refine the triangulation with e.g. Alexander moves, which will lead to another (larger) Hilbert space. The question is to specify so--called embedding or refining maps, that map states from a `coarser' Hilbert space to states in a `finer' Hilbert space \cite{BD12b,BD14}. Such embeddings impose a certain vacuum state \cite{TimeEvol} and allow the construction of a continuum Hilbert space via an inductive limit (if certain consistency conditions are satisfied) \cite{DG}. A particular such embedding map would impose a quantum--deformed BF vacuum: Here one would just surround each new triangulation edge with a vacuum loop. (In the ${\cal B}_1$ basis, in which the strands of the under--crossing graph are parallel to the edges, the BF vacuum associates $j=0$ labels to such strands.) But it seems also possible to construct other refining maps, e.g. corresponding to a (Ashtekar--Lewandowski like) vacuum, in which spatial geometry operators vanish.  Additionally one can use the refinement moves  to impose a (gravitational) dynamics, see below. 

~\\
{\bf Coarse graining:} To construct the continuum limit for the dynamics of a given discrete theory it is helpful to understand coarse graining of the associated state spaces. The fusion basis in $(2+1)$ dimensions is perfectly suited for coarse graining \cite{ABC16a}, as it encodes excitations on different coarse graining scales.  The ${\cal B}_1$ and ${\cal B}_2$ basis describe however excitations on the lattice scale (with respect to two different vacua). By changing the pant decomposition of the Heegard surface, which determines the fusion basis, we can also obtain a basis that describes the fusion or coarse graining of excitations. E.g. start from the curvature basis and contract the Heegard surface of genus $g$ along a spanning tree of the spine. That will deform the surface such that it appears as a $2g$ punctured sphere with $g$ handles glued onto it. The decomposition of the the $2g$ punctured sphere into three--punctured spheres or pants will determine a fusion scheme for the basic curvature excitations, defined by the handles. Likewise one can start from the ${\cal B}_2$ basis, which would define a coarse graining or fusion of flux (or spatial geometry) excitations. 

The coarse graining of non--Abelian gauge theories leads to the emergence of torsion degrees of freedom, which on the lattice level are described by violations of the Gau\ss~constraints \cite{EteraCoarse,DG,ABC16a}.  This brings as to the question of how to accommodate such torsion degrees of freedom, see also the paragraph on the introduction of boundaries. 

Finally one would like to implement a `flow of the cosmological constant'. On the classical level this is realized in \cite{BahrDittrichL}. A different strategy for the quantum strategy would  be to start from the state space for a small cosmological constant (large ${\rm k}$) and to impose curvature constraints with a large cosmological constant (corresponding to a smaller ${\rm k}$). See \cite{Etera16L} for a perturbative ansatz based on path integral quantization. 

~\\
{\bf Geometric interpretation of observables and vacua:} In a quantum gravity context a more in--depth analysis of the various Wilson loop operators is needed, and in particular a geometric interpretation. We argued that Wilson loops around edges and the Wilson loops around triangles are related to holonomy and flux operators respectively. This is also in line with the construction \cite{HHKR}. In addition there are Wilson loop operators around other curves, which might be diagonalized by a basis which is neither a ${\cal B}_1$ and ${\cal B}_2$ basis. What is the  geometrical interpretation of these operators? In addition to Wilson loop operators one can consider grasping operators \cite{SmolinQ}, which usually implement the fluxes, and can be straightforwardly evaluated using (\ref{Fmove},\ref{bubble move}). How are these operators geometrically interpreted \cite{Girelli1,DeformedHyper}? Do they provide an alternative quantization of (possibly non--exponentiated) flux operators?   Conversely can one construct operators for more involved geometric quantities like the three--volume \cite{RovelliSmolin,Vol}?

We provided a set of bases which arose from the fusion bases. Can we re--construct the (quantum) geometries described by these states? The work in \cite{HHKR} is discussing a reconstruction for states in the ${\cal B}_2$ basis. Another  interesting case is the curvature basis.  Is there a deeper meaning to the self--duality between the spin network like basis and the curvature basis? A possible reason could be polar duality \cite{kokkendorff}.   Another possible link is the interpretation of the Gau\ss~constraints (or closure constraints) as Bianchi identity and a possible re--construction of a new kind of connection proposed in \cite{Charles2}.

With each generalized fusion basis we have also an a priori different vacuum state. What is the geometric interpretation of these states? Can these states be interpreted as coherent states? 

~\\
{\bf Including Immirzi-- and theta--parameter:} $(3+1)$--dimensional loop quantum gravity is based on the Ashtekar variables \cite{Ashtekar}, that can be generalized to include an Immirzi--Barbero parameter \cite{IB}. Whereas the state space does not change (as long as the parameter is real) the parameter leads to a re--interpretation of the geometric meaning of the holonomy and flux operators.  What would be the role of such a Immirzi--Barbero parameter in this current framework \cite{DittrichRyan} and how would it influence the geometric interpretation of operators and vacua? See also the recent work \cite{EteraCharles}.

One can furthermore generalize the canonical loop quantum variables to include a theta--parameter \cite{theta}, which would lead to a different geometric re--interpretation of the operators. For the framework presented here, can one relate a change of theta--parameter  to a change of generalized fusion basis? 

Here we choose to impose the 2--handle constraints by inserting over--crossing vacuum loops. We could have also used under--crossing vacuum loops. Is it possible to use a combination of both possibilities?

~\\
{\bf Dynamics:}  Having the state spaces and therefore kinematics of quantum gravity one of course wishes also to implement the dynamics. One possibility is to use Pachner moves, that change the Heegard surface, as suggested in \cite{smolin2}.  These Pachner moves can be interpreted as gluing 4--simplices in different ways to the triangulated hyper--surface, see \cite{Hoehn}, and can thus be implemented by a corresponding gluing of a 4--simplex amplitude, for instance constructed along the lines of \cite{HHKR}. Such Pachner moves should  eventually (that is in the continuum limit) act as refining or coarse graining operators \cite{Hoehn,TimeEvol}. The reason is that time evolution in background independent theories, such as gravity, corresponds to a gauge transformation, namely a diffeomorphism,  and thus should map between (gauge) equivalent states.  

However such a discrete dynamics for four--dimensional gravity, which is implemented (via the 4--simpex gluing) in a local way, breaks four--dimensional diffeomorphism invariance \cite{diffeobreak}. To restore this symmetry one would  have to consider a continuum limit \cite{BahrDittrichL, BDS}, which can be constructed via an auxiliary coarse graining flow \cite{BD14}. For coarse graining schemes along the lines of \cite{TNWCG} the finiteness of the state spaces constructed here facilitates a numerical implementation.


\vspace{1.5cm}

\begin{center}
\textbf{Acknowledgements}
\end{center}
The author would like to thank John Barrett, Cl\'ement Delcamp, Laurent Freidel, Marc Geiller, Etera Livine, Aldo Riello and Lee Smolin for helpful discussions and comments. This work is supported by Perimeter Institute for Theoretical Physics. Research at Perimeter Institute is supported by the Government of Canada through Industry Canada and by the Province of Ontario through the Ministry of Research and Innovation.


\appendix

\section{The quantum group SU(2)${}_\text{k}$}
\label{appendix1}

\noindent 
Here we will give a very short review of some basic facts about  ${\rm SU}(2)_{\rm k}$. A  more extensive introduction can be found in \cite{Qbackground}.
 ${\rm SU}(2)_{\rm k}$, where ${\rm k}$ is a positive integer, can be understood to arise as quantum deformation of $\SU(2)$.  The deformation parameter
\be
q=e^{2\pi\i/({\rm k}+2)}
\ee
is a root of unity. We  define quantum numbers
\be\label{qNumber}
[n]\,:=\,\frac{q^{n/2}-q^{-n/2}}{q^{1/2}-q^{-1/2}}=\frac{\sin\left(\tfrac{\pi}{{\rm k}+2}\,n\right)}{\sin\left(\tfrac{\pi}{{\rm k}+2}\right)},
\q
\forall\,n\in\mathbb{N}-\{0\},
\ee
with $[0]=1$. This leads to quantum dimensions $d_j=[2j+1]$. To avoid negative quantum dimensions (which are associated to so--called trace zero representations), one only allows representations in a range $j\in\{0,1/2,1,\dots,{\rm k}/2\}$. These representations are called admissible. We will frequently encounter the signed quantum dimensions 
\ba
v_j^2\,:=\,(-1)^{2j} d_j
\ea
and their square roots $v_j$ (with $(-1)^{1/2}={\rm i}$).  The total quantum dimension is defined as
\ba
\D&:=&\sqrt{\sum_jv_j^4}=\sqrt{\f{{\rm k}+2}{2}}\sin^{-1}\left(\f{\pi}{{\rm k}+2}\right) \q .
\ea

This brings us to the recoupling theory for ${\rm SU}(2)_{\rm k}$.  As in the group case we can tensor representations (although with a deformed co--product). Admissible triples are triples of representations that include the trivial representation in their tensor product. Such triples are defined by the following conditions:
\be\label{Admissibility}
i\leq j+k,
\q
j\leq i+k,
\q
k\leq i+j,
\q
i+j+k\in\mathbb{N},
\q
i+j+k\leq{\rm k}.
\ee
Note the last condition, which is special to the quantum group ${\rm SU}(2)_{\rm k}$.

The so--called $F$--symbols transform between different bracketings for the tensor product. It also appears in the $F$--move equivalence relation in (\ref{Fmove}). To define the $F$--symbols we introduce first  for any admissible triple $(i,j,k)$ the quantity
\be
\Delta(i,j,k)\,:=\,\delta_{ijk}\sqrt{\f{[i+j-k]![i-j+k]![-i+j+k]!}{[i+j+k+1]!}},
\ee
where $[n]!\coloneqq[n][n-1]\dots[2][1]$.

 The (Racah--Wigner) quantum $\{6j\}$ symbol is then given by the formula
\be
\left\{
\begin{array}{ccc}
i&j&m\\
k&l&n
\end{array}\right\}
:=&\:\Delta(i,j,m)\Delta(i,l,n)\Delta(k,j,n)\Delta(k,l,m)\sum_z(-1)^z[z+1]!\nonumber\\
&\times\f{\Big([i+j+k+l-z]![i+k+m+n-z]![j+l+m+n-z]!\Big)^{-1}}{[z-i-j-m]![z-i-l-n]![z-k-j-n]![z-k-l-m]!},\q
\ee
where the sum runs over
\be
\max(i\!+\!j\!+\!m,i\!+\!l\!+\!n,k\!+\!j\!+\!n,k\!+\!l\!+\!m)\leq z\leq\min(i\!+\!j\!+\!k\!+\!l,i\!+\!k\!+\!m\!+\!n,j\!+\!l\!+\!m\!+\!n).
\ee
Now the $F$-symbols are defined as
\be\label{FDefinition}
F^{ijm}_{kln}
\,:=\,
(-1)^{i+j+k+l} \,\sqrt{[2m+1][2n+1]}\,
\left\{
\begin{array}{ccc}
i&j&m\\
k&l&n
\end{array}\right\}.
\ee

We will furthermore need the so--called $R$--matrices, which allows as to resolve over-- and under crossings in the graphical calculus:
\be\label{braiding}
\pic{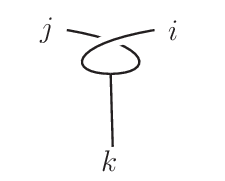}=\,\,R^{ij}_k\pic{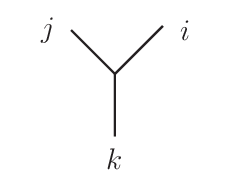},
\q\q
\pic{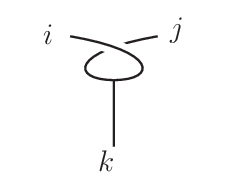}=\,\,\big(R^{ij}_k\big)^*\pic{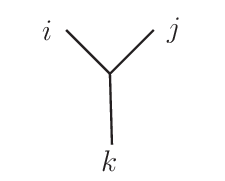}.
\ee
Thus, using a special case of the $F$--move (\ref{Fmove})
\be\label{F split a}
\sum_k\f{v_k}{v_iv_j}\pic{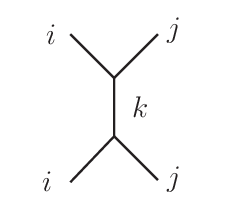}\,=\,\pic{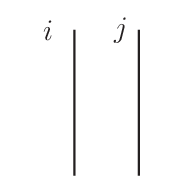},
\ee
 one can conclude that
\be\label{resolution of crossing}
\pic{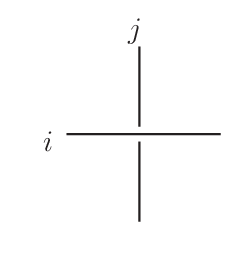}\,=\,\,\sum_k\f{v_k}{v_iv_j}R^{ij}_k\pic{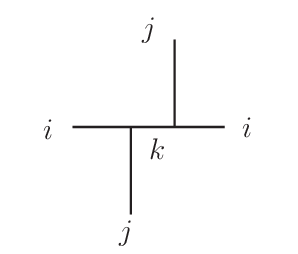},
\,\,
\pic{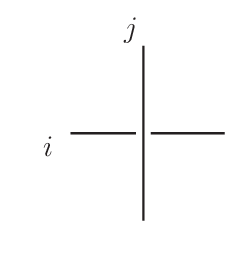}\,=\,\,\sum_k\f{v_k}{v_iv_j}\big(R^{ij}_k\big)^*\pic{Fig/Uncrossing}.
\ee

For ${\rm SU}(2)_{\rm k}$ the $R$-matrix is given as
\ba
R^{ij}_k&=&(-1)^{k-i-j}\,\left(q^{k(k+1)-i(i+1)-j(j+1)}\right)^{1/2}  \q .
\ea

The $S$-matrix is defined as
\be\label{SM1}
\D S_{ij}\,:=\,s_{ij}\,:=\!\pic{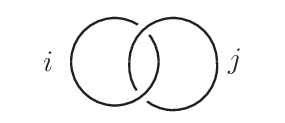}\hs,
\ee
that is one has to remove first the crossings with the help of \eqref{resolution of crossing} and reduce the resulting graphs via $F$--moves to bubbles. These can be related to the empty graph via the bubble move (\ref{bubble move}). The resulting coefficient between the right hand side of (\ref{SM1}) and the empty graph state defines the $s$--matrix:
\be\label{s matrix in terms of R}
s_{ij}\,=\,\sum_l v_l^2R^{ij}_lR^{ji}_l\,=\,(-1)^{2(i+j)}[(2i+1)(2j+1)].
\ee
The $S$-matrix for ${\rm SU}(2)_{\rm k}$ is invertible and unitary, making $SU(2)_{\rm k}$ into a modular fusion category. Note that the $S$--matrix is also real and symmetric:
\be\label{s matrix identities}
S_{ij}=S_{ji},
\q\q
\sum_lS_{il}S_{lj}=\delta_{ij}\q .
\ee

For the basis transformation (\ref{SStrafo})  on the punctured torus  we need   generalizations of the $S$--matrix given by
\ba\label{equAB}
A^i_{kj} \,=\, \frac{1}{{\cal D} \, v_i} \pic{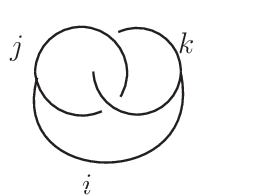} \q ,\q\q    B^i_{kj}\,=\, \frac{1}{{\cal D} \, v_i} \pic{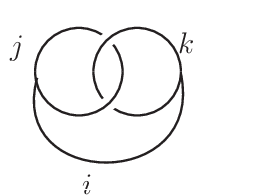} \q .
\ea


\begin{thebibliography}{99}\small

  
  
\bibitem{TV} V. Turaev and O. Viro,
``State sum invariants of 3 manifolds and quantum 6j symbols'',
Topology \textbf{31} 865 (1992).
J. W. Barrett and B. W. Westbury,
``Invariants of piecewise linear three manifolds'',
Trans. Am. Math. Soc. \textbf{348} (1996) 3997, [arXiv:hep-th/9311155]].




\bibitem{BB4D} M.~B\"arenz and J.~Barrett,
  ``Dichromatic state sum models for four-manifolds from pivotal functors,''
  arXiv:1601.03580 [math-ph].
  
  \bibitem{KKR} R. K\"onig, G. Kuperberg and B. W. Reichardt,
``Quantum computation with Turaev--Viro codes'',
Annals of Physics 325, 2707-2749 (2010), arXiv:1002.2816 [quant-ph].

\bibitem{Wu} Y. Hu, N. Geer and Y.-S. Wu,
``Full Dyon Excitation Spectrum in Generalized Levin--Wen Models'',
(2015), arXiv:1502.03433 [cond-mat.str-el].


\bibitem{DG16}   B.~Dittrich and M.~Geiller,
  ``Quantum gravity kinematics from extended TQFTs,'', to appear in NJP, 
  arXiv:1604.05195 [hep-th].
  
  \bibitem{ABC16b}  C.~Delcamp, B.~Dittrich and A.~Riello,
  ``On entanglement entropy in non-Abelian lattice gauge theory and 3D quantum gravity,''
  JHEP {\bf 1611} (2016) 102
  [arXiv:1609.04806 [hep-th]].

  \bibitem{ABC16a}  C.~Delcamp, B.~Dittrich and A.~Riello,
  ``Fusion basis for lattice gauge theory and loop quantum gravity,''
  arXiv:1607.08881 [hep-th].

\bibitem{TimeEvol}
 B.~Dittrich and S.~Steinhaus,
  ``Time evolution as refining, coarse graining and entangling,''
  New J.\ Phys.\  {\bf 16} (2014) 123041
  [arXiv:1311.7565 [gr-qc]].
  
  \bibitem{ALI} A. Ashtekar and C. J. Isham,
``Representations of the holonomy algebras of gravity and non-Abelian gauge theories'',
Class. Quantum Grav. \textbf{9} 1433 (1992), [arXiv:hep-th/9202053].
%
A. Ashtekar and J. Lewandowski,
``Representation theory of analytic holonomy C* algebras'',
in J. Baez, ed., \textit{Knots and Quantum Gravity},
(Oxford University Press, 1994), [arXiv:gr-qc/9311010].
%
A. Ashtekar and J. Lewandowski,
``Projective techniques and functional integration for gauge theories'',
J. Math. Phys. \textbf{36} (1995) 2170, [arXiv:gr-qc/9411046].
%
 A. Ashtekar and J. Lewandowski,
``Differential geometry on the space of connections via graphs and projective limits'',
J. Geom. Phys. \textbf{17} (1995) 191, [arXiv:hep-th/9412073].

\bibitem{BF} G. T. Horowitz,
``Exactly soluble diffeomorphism invariant theories",
Comm. Math. Phys. \textbf{125} 417 (1989).



\bibitem{DG} B. Dittrich and M. Geiller,
``A New Vacuum for Loop Quantum Gravity",
Class. Quantum Grav. \textbf{32} 112001 (2015), [arXiv:1401.6441 [gr-qc]].
 B. Dittrich and M. Geiller,
``Flux formulation of Loop Quantum Gravity: Classical formulation'',
Class. Quantum Grav. \textbf{32} 135016 (2015), [arXiv:1412.3752 [gr-qc]].
 B. Bahr, B. Dittrich and M. Geiller,
``A new realization of quantum geometry'',
(2015), arXiv:1506.08571 [gr-qc].


\bibitem{SmolinQ} S. Major and L. Smolin,
``Quantum deformation of quantum gravity'',
Nucl. Phys. B \textbf{473} (1996) 267, arXiv:gr-qc/9512020.

\bibitem{BahrDittrichL}
B.~Bahr and B.~Dittrich,
  ``Improved and Perfect Actions in Discrete Gravity,''
  Phys.\ Rev.\ D {\bf 80} (2009) 124030
  [arXiv:0907.4323 [gr-qc]].
   B.~Bahr and B.~Dittrich,
  ``Regge calculus from a new angle,''
  New J.\ Phys.\  {\bf 12} (2010) 033010
  [arXiv:0907.4325 [gr-qc]].


\bibitem{Girelli1} M. Dupuis and F. Girelli,
``Observables in Loop Quantum Gravity with a cosmological constant'',
Phys. Rev. D \textbf{90} (2014), \texttt{arXiv:1311.6841 [gr-qc]}.

\bibitem{RV} C.~Rovelli and F.~Vidotto,
  ``Compact phase space, cosmological constant, and discrete time,''
  Phys.\ Rev.\ D {\bf 91} (2015) no.8,  084037
  doi:10.1103/PhysRevD.91.084037
  [arXiv:1502.00278 [gr-qc]].

\bibitem{SFDiv} 
 C.~Perini, C.~Rovelli and S.~Speziale,
  ``Self-energy and vertex radiative corrections in LQG,''
  Phys.\ Lett.\ B {\bf 682} (2009) 78
  [arXiv:0810.1714 [gr-qc]].
 A.~Riello,
  ``Self-energy of the Lorentzian Engle-Pereira-Rovelli-Livine and Freidel-Krasnov model of quantum gravity,''
  Phys.\ Rev.\ D {\bf 88} (2013) no.2,  024011
  [arXiv:1302.1781 [gr-qc]].
  V.~Bonzom and B.~Dittrich,
  ``Bubble divergences and gauge symmetries in spin foams,''
  Phys.\ Rev.\ D {\bf 88} (2013) 124021
  [arXiv:1304.6632 [gr-qc]].
L. Q. Chen,
``Bulk amplitude and degree of divergence in 4d spin foams'',
(2016), arXiv:1602.01825 [gr-qc].

\bibitem{TNWCG}
B.~Dittrich, M.~Martín-Benito and E.~Schnetter,
  ``Coarse graining of spin net models: dynamics of intertwiners,''
  New J.\ Phys.\  {\bf 15} (2013) 103004
  [arXiv:1306.2987 [gr-qc]].
B. Dittrich and W. Kaminski,
``Topological lattice field theories from intertwiner dynamics'',
 arXiv:1311.1798 [gr-qc].
B.~Dittrich, M.~Martin-Benito and S.~Steinhaus,
  ``Quantum group spin nets: refinement limit and relation to spin foams,''
  Phys.\ Rev.\ D {\bf 90} (2014) 024058
  [arXiv:1312.0905 [gr-qc]].
   B.~Dittrich, S.~Mizera and S.~Steinhaus,
  ``Decorated tensor network renormalization for lattice gauge theories and spin foam models,''
  New J.\ Phys.\  {\bf 18} (2016) no.5,  053009
  [arXiv:1409.2407 [gr-qc]].
  S.~Steinhaus,
  ``Coupled intertwiner dynamics: A toy model for coupling matter to spin foam models'',
Phys. Rev. D \textbf{92} (2015) 064007, [arXiv:1506.04749 [gr-qc]].
  B.~Dittrich, E.~Schnetter, C.~J.~Seth and S.~Steinhaus,
 ``Coarse graining flow of spin foam intertwiners,''
  Phys.\ Rev.\ D {\bf 94} (2016) no.12,  124050
  [arXiv:1609.02429 [gr-qc]].
  C.~Delcamp and B.~Dittrich,
  ``Towards a phase diagram for spin foams,''
  arXiv:1612.04506 [gr-qc].



\bibitem{3to4}  C.~Delcamp and B.~Dittrich,
  ``From 3D TQFTs to 4D models with defects,''
  arXiv:1606.02384 [hep-th].
  
  \bibitem{smolin2} F.~Markopoulou and L.~Smolin,
  ``Quantum geometry with intrinsic local causality,''
  Phys.\ Rev.\ D {\bf 58} (1998) 084032
  [gr-qc/9712067].

  
  
  \bibitem{WTR}
   N.~Reshetikhin and V.~G.~Turaev,
  ``Invariants of three manifolds via link polynomials and quantum groups,''
  Invent.\ Math.\  {\bf 103} (1991) 547.
  
  \bibitem{CraneYetter}
 L.~Crane and D.~Yetter,
  ``A Categorical construction of 4-D topological quantum field theories,''
  In *Dayton 1992, Proceedings, Quantum topology* 120-130
  [hep-th/9301062].
  L.~Crane, L.~H.~Kauffman and D.~N.~Yetter,
  ``State sum invariants of four manifolds. 1.,''
  hep-th/9409167.
  
  \bibitem{Barrett4DObs}
 J.~W.~Barrett, J.~M.~Garcia-Islas and J.~F.~Martins,
  ``Observables in the Turaev-Viro and Crane-Yetter models,''
  J.\ Math.\ Phys.\  {\bf 48} (2007) 093508
  [math/0411281 [math.QA]].
  
 
 \bibitem{HHKR} H. M. Haggard, M. Han, W. Kaminski and A. Riello,
``SL(2,C) Chern--Simons Theory, a non-Planar Graph Operator, and 4D Loop Quantum Gravity with a Cosmological Constant: Semiclassical Geometry'',
Nucl. Phys. B \textbf{900}, 1 (2015), [arXiv:1412.7546 [hep-th]].
%
 H. M. Haggard, M. Han and A. Riello,
``Encoding Curved Tetrahedra in Face Holonomies: a Phase Space of Shapes from Group-Valued Moment Maps'',
Annales Henri Poincar\'e 1--48 (2015), [arXiv:1506.03053 [math-ph]].
%
 H.~M.~Haggard, M.~Han, W.~Kaminski and A.~Riello,
  ``Four-dimensional Quantum Gravity with a Cosmological Constant from Three-dimensional Holomorphic Blocks,''
  Phys.\ Lett.\ B {\bf 752} (2016) 258
  [arXiv:1509.00458 [hep-th]].
%
  H.~M.~Haggard, M.~Han, W.~Kaminski and A.~Riello,
  ``SL(2,C) Chern-Simons Theory, Flat Connections, and Four-dimensional Quantum Geometry,''
  arXiv:1512.07690 [hep-th].
 M.~Han and Z.~Huang,
  ``SU(2) Flat Connection on Riemann Surface and Twisted Geometry with Cosmological Constant,''
  arXiv:1610.01246 [gr-qc].
  
  \bibitem{WW}
  K.~Walker and Z.~Wang,
  ``(3+1)-TQFTs and Topological Insulators,''
  arXiv:1104.2632 [cond-mat.str-el].
  
  
  \bibitem{LevinWen} M. A. Levin and X.-G. Wen,
``String-net condensation: A physical mechanism for topological phases'',
Phys. Rev. B \textbf{71} (2005) 045110, [arXiv:cond-mat/0404617].
 T. Lan and X.-G. Wen,
``Topological quasiparticles and the holographic bulk-edge relation in 2+1D string-net models'',
Phys. Rev. B \textbf{90}, 115119 (2014), arXiv:1311.1784 [cond-mat.str-el].

\bibitem{BaezL}
J.~C.~Baez,
  ``Four-Dimensional BF theory with cosmological term as a topological quantum field theory,''
  Lett.\ Math.\ Phys.\  {\bf 38} (1996) 129
  [q-alg/9507006].
  
  
  \bibitem{Burnell}
  C.~W.~von Keyserlingk, F.~J.~Burnell and S.~H.~Simon,
  ``Three-dimensional topological lattice models with surface anyons,''
  Phys.\ Rev.\ B {\bf 87} (2013) no.4,  045107
  [arXiv:1208.5128 [cond-mat.str-el]].
  
  
  \bibitem{BCtoappear}
  C.~Delcamp, B.~Dittrich, ``One fusion basis to rule them all", to appear
  
   \bibitem{RovelliSmolin}
    C.~Rovelli and L.~Smolin,
  ``Discreteness of area and volume in quantum gravity,''
  Nucl.\ Phys.\ B {\bf 442} (1995) 593
   Erratum: [Nucl.\ Phys.\ B {\bf 456} (1995) 753]
  [gr-qc/9411005].
    C.~Rovelli and L.~Smolin,
  ``Spin networks and quantum gravity,''
  Phys.\ Rev.\ D {\bf 52} (1995) 5743
  [gr-qc/9505006].

  \bibitem{BarrettObs}
   J.~W.~Barrett,
  ``Geometrical measurements in three-dimensional quantum gravity,''
  Int.\ J.\ Mod.\ Phys.\ A {\bf 18S2} (2003) 97
  [gr-qc/0203018].
  
  \bibitem{BD12b} B. Dittrich,
``From the discrete to the continuous: Towards a cylindrically consistent dynamics,''
New J. Phys. \textbf{14} (2012) 123004, [arXiv:1205.6127 [gr-qc]].

\bibitem{BD14} B. Dittrich,
``The continuum limit of loop quantum gravity - a framework for solving the theory,''
in A. Ashtekar and J. Pullin, ed., to be published in the World Scientific series ``100 Years of General Relativity'',
(2014), [arXiv:1409.1450 [gr-qc]].
  
  
  \bibitem{Qbackground} A. N. Kirillov and N. Y. Reshetikhin,
``Representations of the algebra Uq (SU(2)), q-orthogonal polynomials and invariants of links,''
in \textit{Infinite dimensional Lie algebras and groups}, ed. by V. G. Kac, pp. 285-339,  World Scientific, Singapore (1989).
 L. H. Kauffman and S. Lins,
\textit{Temperley--Lieb Recoupling Theory and Invariants of 3-Manifolds},
Princeton University Press, (1994).
  J. S. Carter, D. E. Flath and M. Saito,
\textit{The Classical and Quantum 6j-symbols},
Princeton University Press, (1995).

    \bibitem{Kir} A. Kirillov Jr.,
``String-net model of Turaev--Viro invariants,''
(2011), arXiv:1106.6033 [math.AT].

\bibitem{kono} T.~Kohno, ``Topological invariants for 3-manifolds using representations of mapping class groups I," Topology {\bf 31} (1992) 203.

\bibitem{Koenig2} G.~ Alagic, S.~P.~Jordan, R.~Koenig, B.~W.~Reichardt,
``Approximating Turaev-Viro 3-manifold invariants is universal for quantum computation,"
Phys.\ Rev.\ A {\bf 82} (2010) 040302(R) 


\bibitem{Kitaev} A. Y. Kitaev,
``Fault tolerant quantum computation by anyons,''
Annals Phys. \textbf{303} (2003) 2, arXiv:quant-ph/9707021.


\bibitem{HeegardNotes}
J. Johnson, ``Notes on Heegard Splittings," http://users.math.yale.edu/~jj327/notes.pdf

\bibitem{GSbook}
R. E. Gompf and A. Stipsicz, \textit{4-manifolds and Kirby Calculus}, Graduate studies in mathematics, American Mathematical Society, 1999.



\bibitem{BalKir} B. Balsam and A. Kirillov Jr.,
``Turaev--Viro invariants as an extended TQFT'',
(2010), arXiv:1004.1533 [math.GT].
B. Balsam,
``Turaev--Viro invariants as an extended TQFT II'',
(2010), arXiv:1010.1222 [math.QA].
B. Balsam,
``Turaev--Viro invariants as an extended TQFT III'',
(2010), arXiv:1012.0560 [math.QA].

\bibitem{Virelizier} V. Turaev and A. Virelizier,
``On two approaches to 3-dimensional TQFTs'',
(2010), arXiv:1006.3501 [math.GT].

\bibitem{lqg}
 C. Rovelli,
\textit{Quantum Gravity},
(Cambridge University Press, Cambridge, 2004).
T. Thiemann,
\textit{Introduction to Modern Canonical Quantum General Relativity},
(Cambridge University Press, Cambridge, 2007).
A. Perez,
``The Spin Foam Approach to Quantum Gravity'',
Living Rev. Rel. \textbf{16} (2013) 3, [arXiv:1205.2019 [gr-qc]].


\bibitem{Burnell2} F.~J.~Burnell, X.~Chen, L.~Fidkowski, A.~Vishwanath, ``Exactly Soluble Model of a 3D Symmetry Protected Topological Phase of Bosons with Surface Topological Order,"
Phys. Rev. B  {\bf 90} (2014) 245122 [arXiv:1302.7072 [cond-mat.str-el]]

\bibitem{kokkendorff}
S.L.~Kokkendorff,
 ``Polar Duality and the Generalized Law of Sines,''
  Journal of Geometry, {\bf 86} (2007) 140.


\bibitem{dominique} D.~J.~Williamson and Z.~Wang
``Hamiltonian Realizations of (3+1)-TQFTs,''  arXiv:1606.07144 [quant-ph].

\bibitem{smolin1}
 L. Smolin,
``Linking topological quantum field theory and nonperturbative quantum gravity'',
J. Math. Phys. \textbf{36} (1995) 6417, [arXiv:gr-qc/9505028].


\bibitem{BaezPerez}  J.~C.~Baez and A.~Perez,
  ``Quantization of strings and branes coupled to BF theory,''
  Adv.\ Theor.\ Math.\ Phys.\  {\bf 11} (2007) no.3,  451
  [gr-qc/0605087].

\bibitem{EteraCoarse}  E.~R.~Livine,
  ``Deformation Operators of Spin Networks and Coarse-Graining,''
  Class.\ Quant.\ Grav.\  {\bf 31} (2014) 075004
  [arXiv:1310.3362 [gr-qc]].

\bibitem{Etera16L} 
 L.~Freidel and K.~Krasnov,
  ``Spin foam models and the classical action principle,''
  Adv.\ Theor.\ Math.\ Phys.\  {\bf 2} (1999) 1183
  [hep-th/9807092].
L.~Freidel and K.~Krasnov,
  ``Discrete space-time volume for three-dimensional BF theory and quantum gravity,''
  Class.\ Quant.\ Grav.\  {\bf 16} (1999) 351
  [hep-th/9804185].
   E.~R.~Livine,
  ``3d Quantum Gravity: Coarse-Graining and q-Deformation,''
  Ann. Henri Poincar\'e (2016)
  [arXiv:1610.02716 [gr-qc]].

\bibitem{Vol}  A.~Ashtekar and J.~Lewandowski,
  ``Quantum theory of geometry. 2. Volume operators,''
  Adv.\ Theor.\ Math.\ Phys.\  {\bf 1} (1998) 388
  [gr-qc/9711031].

\bibitem{DeformedHyper}
 M.~Dupuis, F.~Girelli and E.~R.~Livine,
  ``Deformed Spinor Networks for Loop Gravity: Towards Hyperbolic Twisted Geometries,''
  Gen.\ Rel.\ Grav.\  {\bf 46} (2014) no.11,  1802
  [arXiv:1403.7482 [gr-qc]].

\bibitem{Charles2} 
 C.~Charles and E.~R.~Livine,
  ``Closure constraints for hyperbolic tetrahedra,''
  Class.\ Quant.\ Grav.\  {\bf 32} (2015) no.13,  135003
  [arXiv:1501.00855 [gr-qc]].
C.~Charles and E.~R.~Livine,
  ``The closure constraint for the hyperbolic tetrahedron as a Bianchi identity,''
  arXiv:1607.08359 [gr-qc].

\bibitem{Ashtekar} 
 A. Ashtekar,
``New variables for classical and quantum gravity'',
Phys. Rev. Lett. \textbf{57}, 2244 (1986).
A. Ashtekar,
``New Hamiltonian formulation of general relativity'',
Phys. Rev. D \textbf{36}, 1587 (1987).

\bibitem{IB} J. F. Barbero,
``Real Ashtekar variables for Lorentzian signature spacetimes'',
Phys. Rev. D \textbf{51} 5507 (1995).
G. Immirzi,
``Real and complex connections for canonical gravity'',
Class. Quant. Grav. \textbf{14} L177 (1997), [arXiv:gr-qc/9612030].
%

\bibitem{DittrichRyan} 
 B.~Dittrich and J.~P.~Ryan,
  ``Phase space descriptions for simplicial 4d geometries,''
  Class.\ Quant.\ Grav.\  {\bf 28} (2011) 065006
  [arXiv:0807.2806 [gr-qc]].
B.~Dittrich and J.~P.~Ryan,
  ``Simplicity in simplicial phase space,''
  Phys.\ Rev.\ D {\bf 82} (2010) 064026
  [arXiv:1006.4295 [gr-qc]].
 B.~Dittrich and J.~P.~Ryan,
  ``On the role of the Barbero-Immirzi parameter in discrete quantum gravity,''
  Class.\ Quant.\ Grav.\  {\bf 30} (2013) 095015
  [arXiv:1209.4892 [gr-qc]].

\bibitem{EteraCharles} C.~Charles and E.~R.~Livine,
  ``Ashtekar-Barbero holonomy on the hyperboloid: Immirzi parameter as a cutoff for quantum gravity,''
  Phys.\ Rev.\ D {\bf 92} (2015) no.12,  124031
  [arXiv:1507.00851 [gr-qc]].

\bibitem{theta} 
A.~Ashtekar, A.~P.~Balachandran and S.~Jo,
  ``The {CP} Problem in Quantum Gravity,''
  Int.\ J.\ Mod.\ Phys.\ A {\bf 4} (1989) 1493.
D.~J.~Rezende and A.~Perez,
  ``The Theta parameter in loop quantum gravity: Effects on quantum geometry and black hole entropy,''
  Phys.\ Rev.\ D {\bf 78} (2008) 084025
  [arXiv:0711.3107 [gr-qc]].
N.~Bodendorfer,
  ``Some notes on the Kodama state, maximal symmetry, and the isolated horizon boundary condition,''
  Phys.\ Rev.\ D {\bf 93} (2016) no.12,  124042
  [arXiv:1602.05499 [gr-qc]].
  
  \bibitem{Hoehn}
   B.~Dittrich and P.~A.~Hohn,
  ``Canonical simplicial gravity,''
  Class.\ Quant.\ Grav.\  {\bf 29} (2012) 115009
  [arXiv:1108.1974 [gr-qc]].
  
 \bibitem{diffeobreak}
 B.~Dittrich,
  ``Diffeomorphism symmetry in quantum gravity models,''
  Adv.\ Sci.\ Lett.\  {\bf 2} 151
  [arXiv:0810.3594 [gr-qc]].
  B.~Bahr and B.~Dittrich,
  ``(Broken) Gauge Symmetries and Constraints in Regge Calculus,''
  Class.\ Quant.\ Grav.\  {\bf 26} (2009) 225011
  [arXiv:0905.1670 [gr-qc]].
 B.~Dittrich and S.~Steinhaus,
  ``Path integral measure and triangulation independence in discrete gravity,''
  Phys.\ Rev.\ D {\bf 85} (2012) 044032
  [arXiv:1110.6866 [gr-qc]].
    B.~Dittrich, W.~Kaminski and S.~Steinhaus,
  ``Discretization independence implies non-locality in 4D discrete quantum gravity,''
  Class.\ Quant.\ Grav.\  {\bf 31} (2014) no.24,  245009
  [arXiv:1404.5288 [gr-qc]].
 
 \bibitem{BDS} 
  B.~Bahr, B.~Dittrich and S.~Steinhaus,
  ``Perfect discretization of reparametrization invariant path integrals,''
  Phys.\ Rev.\ D {\bf 83} (2011) 105026
  [arXiv:1101.4775 [gr-qc]].
   B.~Dittrich,
  ``How to construct diffeomorphism symmetry on the lattice,''
  PoS QGQGS {\bf 2011} (2011) 012
  [arXiv:1201.3840 [gr-qc]].
   B.~Bahr,
  ``On background-independent renormalization of spin foam models,''
  arXiv:1407.7746 [gr-qc].
%

%
\end{thebibliography}
\end{document}